\documentclass[preprint]{aastex}
\usepackage{natbib,graphicx,latexsym}

\newcommand{\HST}{{\sl HST}}
\newcommand{\JWST}{{\sl JWST}}

\newcommand{\Msun}{\mbox{$M_{\sun}$}}

\newcommand{\degree}{\mbox{$^{\circ}$}}

\newcommand{\kms}{\hbox{km~s$^{-1}$}}

\newcommand{\Ks}{\mbox{$K_S$}}

\newcommand{\Mtot}{\mbox{$M_{\rm tot}$}}

\newcommand{\Pcirc}{\mbox{$P_{\rm circ}$}}

\shorttitle{Eccentricities of Very Low-Mass Binaries}
\shortauthors{Dupuy \& Liu}

\begin{document}

\title{On the Distribution of Orbital Eccentricities for Very Low-Mass
  Binaries\altaffilmark{*}}

\author{Trent J.\ Dupuy\altaffilmark{1,2} and
        Michael C.\ Liu\altaffilmark{3}}

      \altaffiltext{*}{The data presented herein were obtained at the
        W.M. Keck Observatory, which is operated as a scientific
        partnership among the California Institute of Technology, the
        University of California, and the National Aeronautics and
        Space Administration. The Observatory was made possible by the
        generous financial support of the W.M. Keck Foundation.}

      \altaffiltext{1}{Harvard-Smithsonian Center for Astrophysics,
        60 Garden Street, Cambridge, MA 02138}

      \altaffiltext{2}{Hubble Fellow}

      \altaffiltext{3}{Institute for Astronomy, University of Hawai`i,
        2680 Woodlawn Drive, Honolulu, HI 96822}

\begin{abstract}

  We have compiled a sample of 16 orbits for very low-mass stellar ($<
  0.1$~\Msun) and brown dwarf binaries, including updated orbits for
  HD~130948BC and LP~415-20AB. This sample enables the first
  comprehensive study of the eccentricity distribution for such
  objects.  We find that very low-mass binaries span a broad range of
  eccentricities from near-circular to highly eccentric ($e \approx
  0.8$), with a median eccentricity of 0.34. We have examined
  potential observational biases in this sample, and for visual
  binaries we show through Monte Carlo simulations that if we choose
  appropriate selection criteria then all eccentricities are equally
  represented ($\lesssim5\%$ difference between input and output
  eccentricity distributions).  The orbits of this sample of very
  low-mass binaries show some significant differences from their
  solar-type counterparts.  They lack a correlation between orbital
  period and eccentricity and display a much higher fraction of
  near-circular orbits ($e<0.1$) than solar-type stars, which together
  may suggest a different formation mechanism or dynamical history for
  these two populations. Very low-mass binaries also do not follow the
  $e^2$ distribution of \citet{ambartsumian37}, which would be
  expected if their orbits were distributed in phase space according
  to a function of energy alone (e.g., the Boltzmann distribution).
  We find that current numerical simulations of very low-mass star
  formation do not completely reproduce the observed properties of our
  binary sample. The cluster formation model of
  \citet{2009MNRAS.392..590B} agrees very well with the overall $e$
  distribution, but the lack of any high-$e$ ($>0.6$) binaries at
  orbital periods comparable to our sample suggests that tidal damping
  due to gas disks may play too large of a role in the simulations. In
  contrast, the circumstellar disk fragmentation model of
  \citet{2009MNRAS.392..413S} predicts only high-$e$ binaries and thus
  is highly inconsistent with our sample. These discrepancies could be
  explained if multiple formation processes are responsible for
  producing the field population.

\end{abstract}



\section{Introduction}

Binary systems have long been used as unique probes into the origins
of stars, as the dynamical imprint of their formation and subsequent
evolution are recorded in their orbits. In fact, decades before it was
known that nucleosynthesis powers stars, well-determined binary orbits
were commonplace \citep[e.g., $>200$ orbits were published in the
compilation of][]{1918QB821.A3a......}. Thus, the study of binary
orbits has a rich heritage, predating much of modern astrophysics.
However, only very recently has it been possible to measure orbits for
the lowest mass products of star formation: stars at the bottom of the
main-sequence ($<0.1$~\Msun) and brown dwarfs.  The sample of orbits
for such objects has been growing rapidly in recent years as
astrometric monitoring has begun to yield orbits for the shortest
period visual binaries known ($P \sim 10$~years), which were
discovered by high-resolution imaging surveys 5--10~years ago
\citep[e.g.,][]{2001AJ....121..489R, 2006AJ....132..891R,
  2003AJ....126.1526B, 2003ApJ...586..512B, 2006ApJS..166..585B,
  2003ApJ...587..407C}.

Some properties of very low-mass binaries, such as orbital semimajor
axes, mass ratios, and multiplicity fractions, can be studied in a
statistical fashion using a large sample without any knowledge of
orbital motion \citep[e.g.,][]{2007ApJ...668..492A,
  2007prpl.conf..427B}.  However, only well-determined orbits can
provide a measure of eccentricity ($e$), which is directly related to
the dynamics of the formation process.  Because of the long history of
stellar orbit studies, many authors have weighed in on the subject of
the eccentricity distribution.\footnote{According to
  \citet{ambartsumian37}, ``the distribution of eccentricities of
  double star orbits has been discussed often enough.''} The most
recent comprehensive work on the eccentricities of solar-type field
binaries was done by \citet{1991A&A...248..485D}, and their
pre--main-sequence counterparts have been reviewed by
\citet{1994ARA&A..32..465M}.  Now for the first time it is possible to
bring similar observations to bear on the extreme low-mass end of the
binary population. Some of the proposed formation mechanisms for such
low-mass objects, such as ejection
\citep[e.g.,][]{2001AJ....122..432R} and gravitational instability
\citep[e.g.,][]{2003MNRAS.346L..36R, 2007MNRAS.382L..30S}, may leave a
distinct signature in the orbits of binaries. Here we present the
first comprehensive look at what orbital eccentricity measurements
reveal about the nature of very low-mass stars and brown dwarfs.


\section{Updated Orbit Determinations \label{sec:orbits}}

We first updated published orbit determinations for very low-mass
binaries that have new data before compiling our sample for this
study.  For two visual binaries (HD~130948BC and LP~415-20AB), we
present new astrometry from our ongoing Keck adaptive optics (AO)
monitoring program in order to refine their orbital elements.  Our
procedure for obtaining, reducing, and analyzing our natural guide
star and laser guide star AO images is described in detail in our
previous work \citep[e.g.,][]{me-130948, me-2397a}, and the binary
parameters derived from our new observations are shown in
Table~\ref{tbl:astrom}.  We used the Markov Chain Monte Carlo (MCMC)
approach described by \citet{liu-2m1534} to determine the posterior
probability distributions of the orbital elements for these binaries,
and in Table~\ref{tbl:orbits} we give their median and 68.3\%
confidence limits.  For HD~130948BC, we combined our new astrometry
with the epochs published by \citet{me-130948}.  We updated this
astrometry to account for the newer astrometric calibration of NIRC2
by \citet{2008ApJ...689.1044G}, which amounts to a $+0\fdg26$ offset
to our previously published position angles.  The extended orbital
phase coverage results in a factor of $\approx$4 improvement in
precision compared to the published orbital parameters.

Some binaries have multiple orbit determinations in the literature,
and in these cases we used our published results in the following
analysis.  In most cases published orbits are in good agreement with
each other, but a few binaries warrant special mention.  In the case
of LP~415-20AB, our astrometry results in a very different orbit than
found by \citet{qk10}. As a test, we fitted their three Keck epochs
with the published Gemini epoch from \citet{2003ApJ...598.1265S} and
found an orbit consistent with the one they reported ($e=0.9\pm0.1$).
However, after including our four Keck epochs we found that the 2007
December astrometry of \citet{qk10} is highly discrepant with the
ensemble of measurements, resulting in a $\chi^2$ of 21.5 for 9
degrees of freedom (DOF). If we exclude this one measurement, we find
good agreement between the data sets, and the best-fit orbit has a
reasonable $\chi^2$ (6.83 for 7 DOF) and significantly different
orbital parameters (e.g., our newly derived eccentricity of
$0.708^{+0.014}_{-0.016}$ is 1.9$\sigma$ discrepant with their
$0.9\pm0.1$). It appears that with only four epochs (i.e., 1 DOF) the
orbit fit of \citet{qk10} was vulnerable to a single discrepant
astrometric measurement leading to an incorrect orbit solution.  We
also note that our published orbit for 2MASS~J2206$-$2047AB
\citep{me-2206} is nominally inconsistent with that published by
\citet{qk10}, with all orbital parameters discrepant by
$\approx$2--4$\sigma$. Using only their data, we have found a lower
$\chi^2$ orbit solution than they report, and although it is more
consistent with our published orbit, it still has an unrealistically
large $\chi^2$ (30.3 for 9 DOF). Thus, we choose to use our published
orbit for 2MASS~J2206$-$2047AB ($\chi^2 = 7.49$ for 7 DOF) in the
following analysis.  Finally, the orbit for 2MASS~J1534$-$2962AB
published by \citet{qk10} is marginally inconsistent with the original
orbit from \citet{liu-2m1534}.  With the latest astrometry the period
and semimajor axis are smaller by 1.8$\sigma$, while the eccentricity
and mass are consistent within 1$\sigma$.  We use the more recent
orbit from \citet{qk10} in the following analysis.


\section{Very Low-Mass Binary Sample \label{sec:sample}}

We have constructed a sample of very low-mass ($\Mtot < 0.2$~\Msun)
binary orbits, including both visual and spectroscopic binaries.
Because it is imperative for our purposes that this sample be as
unbiased as possible with respect to eccentricity, in the following
analysis we consider potential sources of selection bias, the nature
of which varies between different types of binaries. Below we describe
how we have selected our orbit sample as well as the tests we have
performed to quantify the level of eccentricity bias present in this
sample.

\subsection{Visual Binaries \label{sec:vb}}

In Table~\ref{tbl:vb}, we present a list of all 20 visual binaries
that have published orbit determinations and a total system mass
$<0.2$~\Msun. For each binary we list its distance, semimajor axis
($a$), and eccentricity ($e$), as these are the physical parameters
that, along with three viewing angle parameters, determine the size
and shape of a binary's orbit on the sky. Thus, these are the three
key parameters that must be considered when characterizing selection
biases. In most cases we have used orbit parameters as published and
distances determined from trigonometric parallax measurements. When
parallax measurements were not available, we estimated distances based
on the primary components' resolved photometry and spectral types
using the updated spectral type--absolute magnitude relations
described in Appendix~B of \citet{liu-1209}.

For all orbits published by other authors, we have created our own
Markov chains using the published astrometry and the same analysis
method described in Section~\ref{sec:orbits}. However, only in the
cases where the astrometric orbit fits yield a reasonable $\chi^2$ can
we trust the MCMC analysis to provide accurate posterior distributions
of the orbital parameters. In cases where $\chi^2$ is unreasonably
large, the parameter space is not fully explored and uncertainties can
be significantly underestimated by the MCMC method (or in fact by any
fitting method). We find such large $\chi^2$ values (reduced $\chi^2 =
2.8$--120) for five of the orbits published by \citet{qk10}. For three
of these orbits they combined astrometry with radial velocities
(2MASS~J1847$+$5522AB, 2MASS~J1750$+$4424AB, 2MASS~J1426$+$1557AB),
which may have enabled them to achieve orbit fits with smaller
$\chi^2$ than we found using the astrometry alone. For the other two
orbits (2MASS~J0850$+$1057AB and 2MASS~J1728$+$3948AB), it is unclear
why we are unable to reproduce their lower $\chi^2$ values from the
same input data (astrometry alone). For all five of these orbits we
use the parameters derived by \citet{qk10}. (Note that for reasons
unrelated these discrepancies, none of these five binaries ultimately
passes the selection criteria that we define below for our unbiased
eccentricity sample.)

\subsubsection{Selecting a Minimally Biased Sample \label{sec:bias}}

With a complete set of the known very low-mass visual binary orbits in
hand, we now consider how selection biases may affect the eccentricity
distribution of this sample.  Visual binaries must pass three
independent selection criteria before having their orbits determined:
(1) they must be initially resolved by a high-resolution imaging
survey; (2) they must be included in subsequent astrometric monitoring
programs; and (3) their observed relative motion must yield a unique
orbit fit. The first selection cut can result in biases on orbital
parameters, as has been shown for direct imaging searches for
extrasolar planets (e.g., see \citealp{2004ApJ...607.1003B}).  The
second cut results from the fact that it is impractical to monitor the
orbits of all $\approx$100 known very low-mass binaries.\footnote{See
  \texttt{www.vlmbinaries.org}, last updated 2009 July 28.} Rather,
astrometric monitoring programs are inevitably optimized for measuring
as many orbits as possible with limited observing resources over a
limited amount of time by focusing on the binaries likely to have the
shortest orbital periods. Thus, the selection criteria used to define
the monitoring sample can result in strong biases on the orbit
parameters. Because the orbital period must be estimated from the
projected separation of the binary, binaries with wide separations are
much less desirable to monitor. This introduces an eccentricity bias
because more eccentric binaries at a given period and mass will, on
average, have wider projected separations. We will therefore attempt
to define a subset of visual binaries that has minimal eccentricity
bias due to these selection criteria.

\begin{figure}

\centerline{\includegraphics[width=3.0in,angle=0]{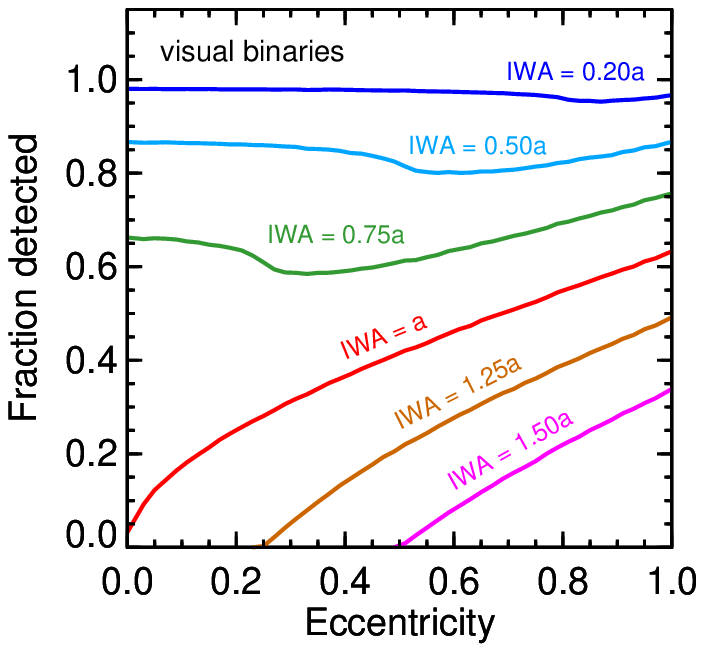}}
\vskip -0.2in
\centerline{\includegraphics[width=3.0in,angle=0]{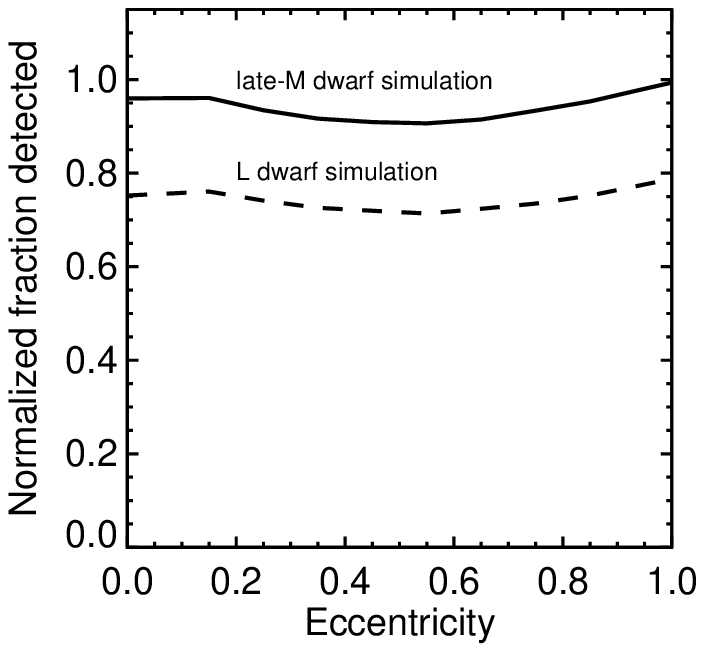}}

\caption{\emph{Top:} Fraction of visual binaries detected in Monte
  Carlo simulations of a single-epoch imaging survey, where the input
  eccentricity distribution is uniform. Different colored lines show
  simulations in which the semimajor axis ($a$) was fixed at different
  multiples of the resolution limit, or inner working angle (IWA).
  When the semimajor axis lies inside the IWA highly eccentric orbits
  are strongly preferred, and as the IWA goes to zero the eccentricity
  bias disappears.  \emph{Bottom:} Eccentricity bias for binaries with
  late-M (solid) and L dwarf (dashed) primaries as determined from
  Monte Carlo simulations assuming a magnitude-limited sample with
  semimajor axis drawn from the log-normal \citet{2007ApJ...668..492A}
  distribution.  The resolution limit for both simulations was chosen
  to be the worst-case from among all surveys that discovered the
  binaries in our sample (0$\farcs$12 for late-M dwarfs and
  0$\farcs$07 for L dwarfs).  The resulting worst-case eccentricity
  bias is modest, only slightly ($\approx$5\%) favoring the discovery
  of highly eccentric ($e>0.8$) and nearly circular ($e<0.2$)
  binaries. \label{fig:bias}}

\end{figure}

{\bf $\bullet$ Discovery bias.} We first consider the effect of
``discovery bias'' which causes the population of binaries resolved by
imaging surveys to deviate from the true underlying eccentricity
distribution of all binaries.  For a given imaging surveys, the extent
of this bias depends on how the semimajor axis ($a$) distribution
compares to that survey's resolution limit, or inner working angle
(IWA).  For example, if the $a$ distribution peaks inside the IWA, the
discovery of higher eccentricity binaries is favored, as such binaries
spend more time at wider separations compared to more circular orbits.
To quantify this bias, we have performed a simple Monte Carlo
simulation in which $a$ was held fixed, while eccentricity was
distributed uniformly from 0 to 1, viewing angles were drawn from
appropriate random distributions ($p(i) \propto \sin{i}$; uniform in
$\omega$ and $\Omega$), and the single-epoch observations were
distributed uniformly in orbital phase.  Binaries were considered
detected if their projected separation at the time of observation was
above the resolution limit.  The impact on the uniform input
eccentricity distribution is illustrated in Figure~\ref{fig:bias}.
When the resolution limit falls at or outside the fixed semimajor axis
(IWA~$\geq~a$), high-$e$ orbits are strongly preferred as they spend
the most time wider than the resolution limit. When the resolution
limit falls inside the fixed value of $a$ (IWA~$<a$) both highly
eccentric and circular orbits are preferred, with intermediate
eccentricities being selected against.  This is because in these cases
any sufficiently eccentric orbit can be made unresolvable by an
unfavorable viewing angle, but this can be compensated by the fact
that the most eccentric orbits spend extended periods of time at very
wide separations. Thus, high-$e$ orbits are more likely to be detected
than intermediate-$e$ orbits, as they spend the most time widely
separated.

To predict the exact form of the eccentricity discovery bias for our
sample would require an assumption for the underlying semimajor axis
distribution as well as careful accounting of the resolution limits of
the numerous imaging surveys that have contributed to the current
sample of very low-mass binaries.  However, for our purposes we simply
want to ensure that our sample of orbits has an acceptably small
amount of bias.  Thus, we need only chose a cutoff in the ratio of
IWA/$a$ that is small enough to minimize the bias while including as
many binaries with orbit determinations as possible.  In
Table~\ref{tbl:vb}, we give our estimate for the IWA of the imaging
data used to discover each visual binary as well as the ratio IWA/$a$.
We have chosen a cutoff of IWA/$a < 0.75$, which Figure~\ref{fig:bias}
shows produces a relatively unbiased selection over a full range of
eccentricities from 0 to 1.  This cutoff excludes three binaries:
LP~415-20AB (IWA/$a \approx 1.28$), 2MASS~J0920$+$3517AB (IWA/$a
\approx 0.91$), and 2MASS~J2140$+$1625AB (IWA/$a \approx 0.83$).  The
highest remaining IWA/$a$ value in our sample is 0.68 for LP~349-25AB.
While it is unfortunate to have to exclude any orbits from our sample,
we emphasize that this is a very important step in keeping potentially
large biases from impacting the following analysis.  The case of
LP~415-20AB is a prime example of the problem with discovery bias on
eccentricity.  This binary was discovered at a projected separation of
$119\pm8$~mas, right at the resolution limit of the imaging survey of
\citet{2003ApJ...598.1265S}.  Its actual semimajor axis turns out to
be $93.5^{+3.1}_{-2.3}$~mas, so it could have only been discovered by
that survey if it had an eccentric orbit, as it indeed does ($e =
0.708^{+0.016}_{-0.014}$).  Thus, while LP~415-20AB does tell us that
very low-mass binaries can be quite eccentric, it cannot convey any
other information about the eccentricity distribution since the only
reason it was discovered is because of its large eccentricity.

{\bf $\bullet$ Monitoring sample bias.} Next, we consider how to
prevent the sample selection of orbital monitoring programs from
biasing the resulting eccentricity distribution. This is essentially a
matter of choosing a range of semimajor axes for which the monitoring
sample is complete at all eccentricities. The lower limit in $a$ is
simply defined by the visual binary orbit with the smallest semimajor
axis; this is Gl~569Bab with $a=0.923\pm0.018$~AU
\citep{me-latem}. There are also no known visual binaries with
projected separations ($\rho$) at discovery smaller than 0.9~AU, so we
define this as the lower limit of our sample. The upper limit should
be chosen to accurately reflect the completeness of orbit monitoring
surveys that tend to focus on the smallest separation binaries, while
also including as many of the resulting orbit measurements as
possible. For the many binaries without orbit determinations, we must
use their minimum possible semimajor axis, which would be the case of
a highly eccentric, face-on orbit caught at apoastron (physical
separation of $a(1+e)$).  Since bound orbits have $0<e<1$, the minimum
possible semimajor axis for any binary is therefore half its projected
separation ($\rho/2$). Thus, some binaries that seem to be very wide
(i.e., poor monitoring targets) could in fact have much smaller
semimajor axes but just be very eccentric. When choosing an upper
limit in $a$, we must therefore ensure that orbit monitoring programs
are complete up to projected separations of $2a$, otherwise we would
introduce a strong selection bias against high eccentricities. We
choose an upper limit in $a$ of 3.7~AU in order to include the
binaries LHS~1070BC ($3.57\pm0.07$~AU; \citealp{2008A&A...484..429S})
and LHS~1901AB ($3.70\pm0.16$~AU; \citealp{me-latem}). The next widest
binary that would also pass our IWA/$a$ cut is 2MASS~0850$+$1057AB
($4.8^{+3.9}_{-1.4}$~AU, $e=0.64\pm0.26$).  Our chosen 3.7~AU upper
limit means that binaries up to projected separations of 7.4~AU should
not have been intentionally excluded from orbit monitoring programs.
To the best of our knowledge this is the case, as at least a second
epoch has been obtained for all binaries with $\rho < 7.4$~AU, either
by our own Keck program or others \citep[e.g.,][]{2008A&A...481..757B,
  qk10}. A slightly larger upper limit of 4.0~AU would require several
more binaries to have undergone orbit monitoring that have not, while
adding no additional binaries with orbit determinations to our sample.

{\bf $\bullet$ Monte Carlo simulations.} With these selection criteria
of $0.9 < a < 3.7$~AU, $\rho < 7.4$~AU, and IWA$/a < 0.75$ in hand, we
have performed Monte Carlo simulations to assess the level of
eccentricity bias that is expected for our sample. For this range of
semimajor axes, the very low-mass binary parameter distributions
derived by \citet{2007ApJ...668..492A} are appropriate. The semimajor
axis distribution is lognormal, with a peak at $\log(a/{\rm AU}) =
0.86$ and $\sigma_{\log{a}} = 0.28$. To convert $a$ from AU to
arcseconds, we randomly drew distances in a Monte Carlo fashion
incorporating the selection effects of the seeing-limited surveys that
originally discovered the binaries as unresolved sources.  We began
with objects distributed uniformly in volume. We assigned magnitudes
to these objects according to the empirical \Ks-band luminosity
function of \citet{2007AJ....133..439C} which is valid from $9.5 <
M_{\Ks} < 13.0$~mag (i.e., $\approx$M8 to L8, encompassing the
spectral type range of the bulk of our sample).  We further added flux
due to unresolved companions according to the
\citet{2007ApJ...668..492A} binary frequency of 20\% and mass ratio
distribution of $q^{1.8}$; we converted mass ratio to flux ratio using
the $L \propto M^{2.4}$ theoretical power-law relation of
\citet{bur01}. The final distance distribution was then determined by
magnitude cuts on this simulated population. We tested both strong
($\Ks < 12.0$~mag) and weak ($\Ks < 15.0$~mag) magnitude cuts
equivalent to those used by \citet{2000AJ....120.1085G} and
\citet{2000AJ....120..447K} in mining 2MASS for late-M dwarfs and L
dwarfs, respectively. In the latter case, we applied an additional
$M_{\Ks} > 11.0$~mag cut to approximate the $J-K > 1.7$~mag color cut
of \citet{2000AJ....120..447K} that excluded brighter, bluer M dwarfs;
we based this cut on the color--absolute magnitude diagrams of
\citet{2002AJ....124.1170D}.

The strong magnitude cut for the brighter late-M sample resulted in a
distance distribution peaking at $\approx$20~pc, while the weaker cut
for the fainter L dwarf sample resulted in similarly shaped
distribution peaked at $\approx$40~pc. We then randomly generated an
observed projected separation ($\rho$) for each binary in a similar
manner as described earlier (i.e., random viewing angles and observing
times, and a uniform eccentricity distribution). Figure~\ref{fig:bias}
shows the resulting eccentricity distribution after applying the
selection criteria of $0.9 < a < 3.7$~AU, IWA$/a < 0.75$, $\rho <
7.4$~AU, and $\rho > {\rm IWA}$ (i.e., the binary was resolved). For
both samples we conservatively assumed the worst-case IWA from the
surveys that originally discovered the binaries in our sample. For the
late-M dwarfs this is 0$\farcs$12 from the Gemini AO surveys
\citep[e.g.,][]{2002ApJ...567L..53C, 2003ApJ...587..407C,
  2003ApJ...598.1265S}, and for the L dwarfs this is 0$\farcs$07 from
both \HST\ and Keck surveys \citep[e.g.,][]{2003AJ....126.1526B,
  2006AJ....132..891R, 2007AJ....133.2320S}. By assuming the
worst-case IWA we find the largest possible amplitude of the
eccentricity bias, since smaller values of the IWA will always yield
less bias (i.e., more binaries will always be uncovered for a smaller
IWA). The resulting maximal bias we find from our Monte Carlo
simulations has an amplitude of less than $\pm5\%$ for both the late-M
and L samples. As expected, the level of eccentricity bias introduced
by our selection criteria is negligible, since they were chosen to
minimize the eccentricity bias, and as shown in Figure~\ref{fig:bias}
there is no strong systematic trend in the bias.  Nearly circular
($e<0.2$) and highly eccentric orbits ($e>0.8$) are slightly preferred
(by $\approx$4\%) as compared to intermediate eccentricities.  We
neglect this bias in the following analysis.

{\bf $\bullet$ Orbit fitting bias.} Finally, we consider one additional
selection effect present in our sample: each binary must have
undergone astrometric motion suitable for yielding a robust orbit
determination. This can potentially introduce an eccentricity bias if
orbit fits are more easily determined for certain ranges of
eccentricity.  For example, it is easy to imagine that an extremely
eccentric binary would present a challenge to astrometric monitoring
as it spends most of its time at a fixed projected separation before
moving rapidly inward in nearly a straight line.  Such an orbit would
be difficult to observe at just the right times needed for a good fit
and would also appear to be less appealing for continued monitoring.
\citet{1977PASP...89..400H} investigated this bias by generating
$10^3$ random orbits and classifying the expected quality of the orbit
fit by eye, relying on their extensive experience working with such
astrometric data. Such an ad hoc method is necessary because in the
marginal cases of visual binary orbit fitting the process of $\chi^2$
minimization is highly nonlinear and strongly dependent on the initial
guess. \citet{1977PASP...89..400H} found that on average orbits with
$e < 0.6$ all had the same likelihood of yielding a good fit, so they
assigned this eccentricity range a normalized completeness factor of
100\%. They found that this completeness drops to 77\% for
eccentricities of 0.6--0.8 and 42\% for $e > 0.8$. In our sample, we
have one visual binary with $e>0.8$ and none in the 0.6--0.8 range. In
the following analysis we consider the impact of applying correction
factors of 1.3$\times$ and 2.4$\times$ due to this orbit fitting bias.

{\bf $\bullet$ Final visual binary sample.} In Table~\ref{tbl:ecc}, we
list the orbital periods, eccentricities, and masses of the eleven
visual binary orbits from Table~\ref{tbl:vb} that meet our
eccentricity bias-minimizing criteria of having $0.9 < a < 3.7$~AU and
a discovery IWA$/a < 0.75$. (Note that we list orbital period here
rather than semimajor axis in order to compare these orbits directly
to the unresolved spectroscopic binary orbits.)  Although this
nominally includes only a little over half of the 20 published visual
binary orbit determinations, in fact the majority of the remaining
orbits would not contribute much information as they are not as well
constrained (see the orbit quality metrics given in
Table~\ref{tbl:vb}).  For example, all orbits in our bias-minimized
sample happen to have 1$\sigma$ eccentricity confidence limits of
$\lesssim0.1$ (median 0.005).  However, of the nine orbits not
included in the sample, six of them have much more poorly constrained
orbits (1$\sigma$ confidence limits of 0.1--0.5), with a median error
of 0.26 that is $>50\times$ larger than the median for the other
orbits.  In fact, the reason why these six orbits are poorly
determined is the same as why they were excluded by our bias
minimizing criteria: they are all the widest (i.e., longest period)
binaries, and this results in them having orbit monitoring
observations that cover much less orbital phase (as little as 0.2\%
and 1.3\% of the orbit is covered in the two most extreme cases).

\subsection{Spectroscopic Binaries \label{sec:sb}}

To date, five unresolved very low-mass binaries have published orbit
determinations. Two of these were discovered via radial velocity (RV)
monitoring as single-lined spectroscopic binaries (SB1s):
Cha~H$\alpha$~8 \citep[$e=0.59\pm0.22$;][]{2007ApJ...666L.113J}, and
2MASS~J0320$-$0446 \citep[$e=0.067\pm0.015$;][]{2008ApJ...678L.125B}.
\citet{2008ApJ...685..553S} have examined the selection bias
introduced by RV surveys on the eccentricity distribution through
Monte Carlo simulations.  They found that as long as the RV
semiamplitude ($K_1$) of an SB1 is $\gtrsim$3$\times$ larger than the
rms of the residuals ($\sigma$), its eccentricity does not greatly
impact its likelihood of discovery.  As shown in their Figure~4, SB1s
have a constant detection efficiency up to $e=0.6$, beyond which there
is a slight downturn.  However, even at the highest eccentricities
($0.85 < e < 1.0$) the detection efficiency is only $\approx$20\% less
than for $e<0.6$ orbits.  The reason for this lack of a strong
eccentricity bias in RV data is because eccentricity acts to boost the
RV semiamplitude ($K_1 \propto 1/\sqrt{1-e^2}$), counteracting the
fact that eccentric orbits are more difficult to adequately sample
with times series data.  The two SB1s in our sample, Cha~H$\alpha$~8
and 2MASS~J0320$-$0446, have large semiamplitudes relative to their
rms ($K_1/\sigma = 11$ and 21, respectively), and thus are expected to
harbor little bias ($<20\%$) with respect to eccentricity according to
the analysis of \citet{2008ApJ...685..553S}.  For this reason and the
small size of the sample (2 systems) we neglect the eccentricity bias
for SB1s.

\begin{figure}

\centerline{\includegraphics[width=3.0in,angle=0]{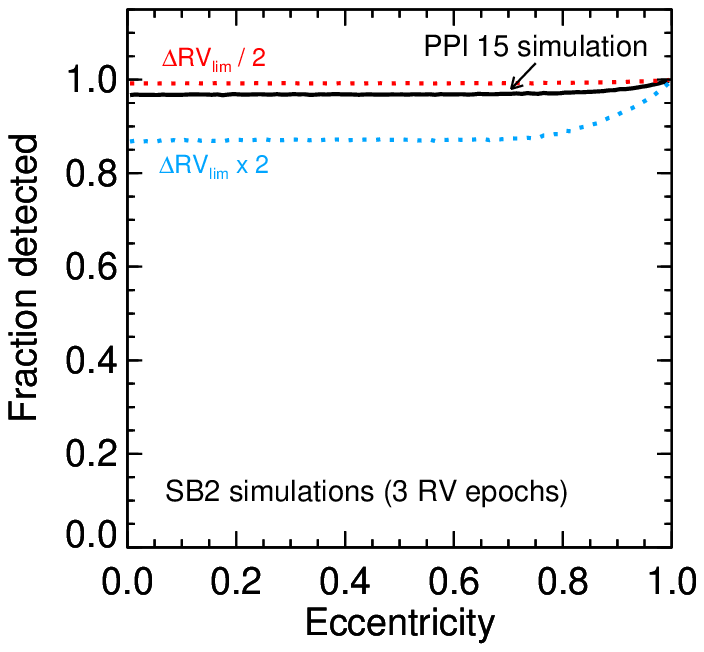}}
\vskip -0.2in
\centerline{\includegraphics[width=3.0in,angle=0]{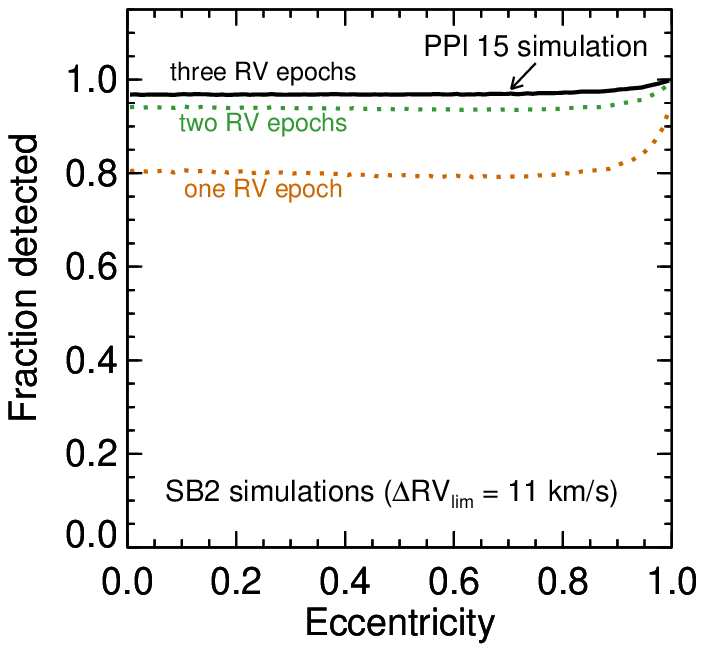}}

\caption{\emph{Top:} Fraction of SB2s detected in Monte Carlo
  simulations of a three-epoch RV survey, where the input eccentricity
  distribution is uniform. The solid black line is a simulation
  designed to imitate the case of PPl~15, the only SB2 in our orbit
  sample.  Dotted colored lines show simulations in which the
  detection threshold ($\Delta{\rm RV} = 11$~\kms) was changed by a
  factor of two.  The RV observations that discovered PPl~15 only
  slightly ($\approx$3\%) favored the detection of highly eccentric
  ($e \gtrsim 0.8$) binaries. This effect would be larger
  ($\approx$15\%) for binaries with a factor of two lower RV
  semiamplitudes.  \emph{Bottom:} Same as top panel, except that the
  dotted colored lines show variations in the eccentricity bias due to
  having fewer than the three epochs of RV data that were available
  for the detection of PPl~15.  Surveys with only one or two RV epochs
  are less sensitive to SB2s at all eccentricities, and they are more
  biased toward the discovery of highly eccentric
  binaries. \label{fig:bias-sb2}}

\end{figure}

One double-lined spectroscopic binary (SB2) has a measured orbit:
PPl~15 \citep[$e=0.42\pm0.05$][]{1999AJ....118.2460B}. It was
discovered as an SB2 in three epochs of RV data. The identification of
SB2s in spectroscopic data is subject to different eccentricity
selection effects than for SB1s because changes in the velocity do not
need to be measured to detect the two sets of absorption lines due to
the binary. To quantify the eccentricity bias in detecting SB2s, we
performed a Monte Carlo simulation similar to the one described above
for visual binaries but with the detection criterion that the velocity
shift between the binary components must be large enough to detect in
high-resolution spectra at one of three random epochs. We fixed the
period and semimajor axis to that of PPl~15 (5.8~days,
0.03~AU).\footnote{Note that our simulation on the detection of SB2s
  is nominally independent of an assumed mass ratio, as $\Delta{\rm
    RV}$ does not depend on this quantity ($\Delta{\rm RV} \propto a
  \sin{i} P^{-1} \sqrt{1-e^2}^{-1}$). Of course, very low mass ratio
  systems of all eccentricities are equally selected against as the
  secondary's absorption lines are more difficult to detect in the
  integrated-light spectrum.} Figure~\ref{fig:bias-sb2} shows the
resulting SB2 eccentricity bias predicted by our simulation for a
detection threshold of $\Delta{\rm RV} = 11$~\kms\ (i.e., 3$\times$
the spectral resolution of \citealp{1999AJ....118.2460B}). We find
only a slight ($\approx$3\%) preference for the discovery of highly
eccentric ($e > 0.8$) SB2s in such RV data. For comparison, we also
show the results if the detection threshold were higher or lower by a
factor of two, as would be the case for binaries with RV
semiamplitudes different by a factor of two. Higher semiamplitude
binaries are less biased, while for lower semiamplitudes the
eccentricity bias becomes a larger effect (though still $\lesssim
15\%$). Thus, it seems reasonable to neglect eccentricity bias for
PPl~15, the one SB2 in our orbit sample.

LSR~J1610$-$0040 ($e=0.444\pm0.017$) is the only very low-mass
astrometric binary known to date, discovered by the USNO parallax
program \citep{2008ApJ...686..548D}. It was later found to be an SB1
in the multi-epoch RV data obtained by \citet{2010ApJ...723..684B}.
In principle, astrometric time series data is very similar to RV time
series data and so would be subject to the same biases as for
SB1s. Thus, we assume that the same selection effects apply, namely
that orbits with $e > 0.85$ have an $\approx$80\% detection efficiency
relative to less eccentric orbits, which are all equally accessible by
astrometric monitoring.

Finally, just one very low-mass eclipsing binary is known,
2MASS~J0535$-$0546 ($e=0.323\pm0.006$), discovered by
\citet{2006Natur.440..311S} in a photometric variability survey. In a
study of exoplanet transit surveys, \citet{2008ApJ...679.1566B} showed
that photometric transit/eclipse detections are only slightly biased
($<10\%$) in eccentricity, with the exact correction depending on the
underlying $e$ distribution. Thus, we neglect any eccentricity bias
for 2MASS~J0535$-$0546.

Table~\ref{tbl:ecc} lists the orbital periods, eccentricities, and
masses for these five unresolved binaries' orbits. Combined with the
visual binary sample, our sample comprises a total of 16 very low-mass
binary orbits spanning orbital periods from 5.8~days to 8400~days.


\section{Discussion}

\begin{figure}

\centerline{
\includegraphics[width=3.2in,angle=0]{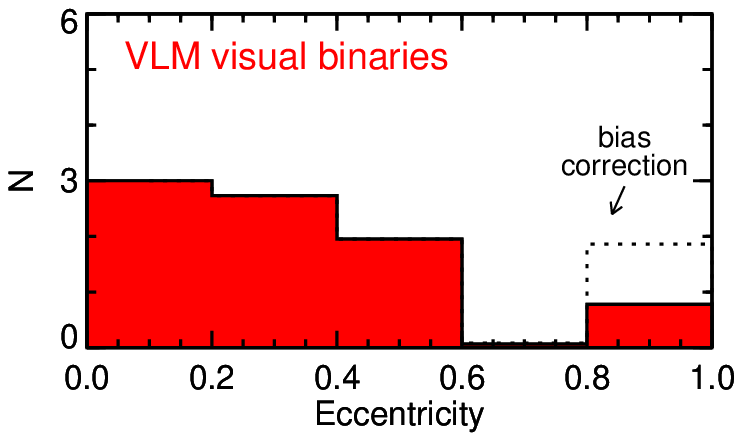}
\hskip 0.2in
\includegraphics[width=3.2in,angle=0]{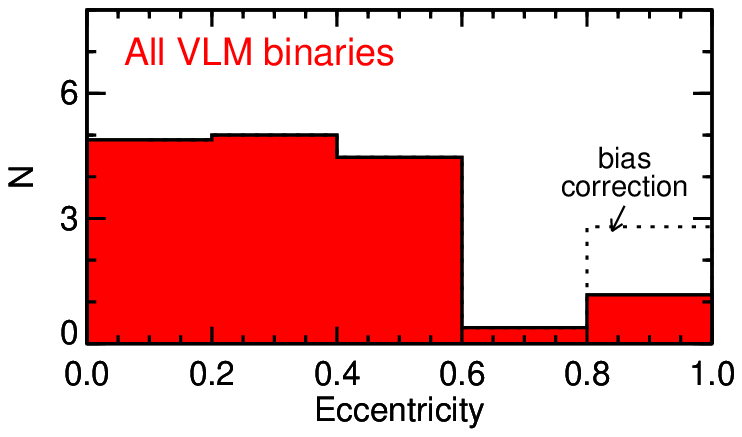}}
\vskip -0.25in
\centerline{
\includegraphics[width=3.2in,angle=0]{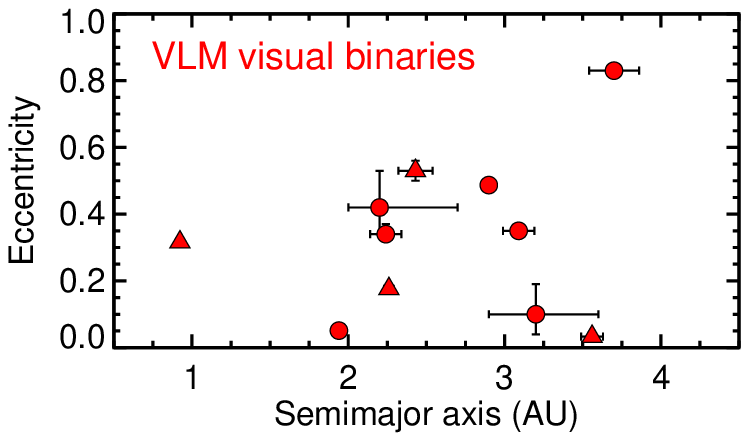}
\hskip 0.2in
\includegraphics[width=3.2in,angle=0]{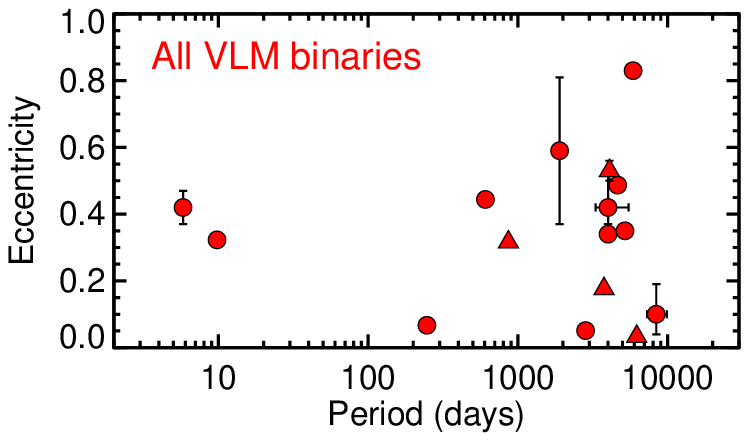}}

\caption{ \emph{Top:} Eccentricity distributions for our sample of
  eleven very low-mass visual binaries (left) and the full sample
  including spectroscopic binaries (right). The dotted histograms show
  the orbit-fitting bias correction we applied to visual binaries in
  the last two bins ($e > 0.6$).  Histograms were generated from
  $10^4$ randomly drawn eccentricity values for each binary
  corresponding to the measurement error, so some bins may have a
  non-integer value.  \emph{Bottom:} Eccentricity plotted as a
  function of semimajor axis for the visual binary sample (left) and
  as a function of orbital period for the full sample including
  spectroscopic binaries (right). Triangles denote members of
  hierarchical triple systems, and circles denote true
  binaries.  \label{fig:vlm}}

\end{figure}

The eccentricity distribution and orbital period--eccentricity diagram
for our sample is shown in Figure~\ref{fig:vlm}. Our sample shows a
strong preference for modest eccentricities, with a median
eccentricity of 0.34 (mean also 0.34) for the full sample and a median
of 0.33 (mean of 0.33) for the visual binaries alone. The visual
binary orbits show a broad range of eccentricities, from near-circular
up to $e = 0.830$. At the shortest periods ($P \lesssim 1000$~days),
we find an absence of eccentric orbits ($e > 0.5$), although this
could be due to small number statistics. Thus, we find no strong
evidence that the eccentricity distribution of very low-mass binaries
changes over three orders of magnitude in orbital period
($6<P<8\times10^3$~days).

We have quantified the lack of a correlation between period and
eccentricity by computing Spearman's rank correlation coefficient
($r_S$) for our sample.  We use a Monte Carlo approach, accounting for
the period and eccentricity errors by drawing the data in each trial
from the observed distributions (our MCMC chains when possible,
Gaussians otherwise).  The resulting mean and rms of the correlation
coefficient was $r_S = -0.02\pm0.07$, which is well below the critical
value of 0.412 needed to show correlation for our sample even at 90\%
confidence. The subset of visual binaries also shows no significant
$P$--$e$ correlation ($r_S = 0.02\pm0.08$, critical value 0.497 at
90\% confidence) or $a$--$e$ correlation ($r_S = 0.14\pm0.09$). In
contrast, \citet{qk10} reported a trend in $e$ as a function of period
for their sample of fifteen very low-mass visual binaries, although
they did not quantify the significance of the correlation. We note
that their observed trend is largely dependent on the two longest
period orbits in their sample ($P = 320\pm240$~yr and
$2000^{+2100}_{-1900}$~yr) for which they found large eccentricities
with large errors ($e = 0.71\pm0.18$ and $0.85^{+0.10}_{-0.41}$,
respectively). Both of these binaries were excluded from our sample by
our criteria for minimizing eccentricity bias, and we also note that
with only 2.2\% and 0.4\% observational phase coverage these orbit
determinations and their uncertainties are not likely to be robust
(e.g., see the cautionary example of DF~Tau~AB in
\citealp{2006AJ....132.2618S}).

We have also investigated whether mass ratio is correlated with
eccentricity for our sample.  In Table~\ref{tbl:ecc}, we list the best
available mass ratio ($q \equiv M_2/M_1$) estimates or measurements
for our sample binaries.  For the visual binaries, $q$ is estimated
from Lyon evolutionary models \citep{2000ApJ...542..464C,
  2003A&A...402..701B} given the measured total mass and individual
luminosities using the method described in Section~5.4 of
\citet{me-130948}. This has been done in a consistent fashion for the
sample presented here, so even if there are systematic errors in the
evolutionary models the relative ordering of the mass ratios should be
relatively unaffected.\footnote{We note that a few visual binaries in
  our sample have mass ratios reported by \citet{qk10} based on
  resolved radial velocities.  However, they have unacceptably large
  uncertainties, with 1$\sigma$ ranges typically spanning a factor of
  $\gtrsim5$ in mass ratio except for Gl~569Bab ($q =
  0.71^{+0.19}_{-0.13}$, i.e., 0.58--0.90 at 1$\sigma$), which is
  still very large for our purposes.  Thus, we have used our
  model-based $q$ estimates for these binaries.}  For the eclipsing
binary 2MASS~J0535$-$0546, we used its directly measured mass ratio
\citep[$0.64\pm0.04$;][]{2006Natur.440..311S}.  None of the three SB1s
in our sample has a well-determined mass ratio, so we excluded them
from our mass ratio analysis.  We also excluded one visual binary
(LHS~2397aAB) because it has a very large uncertainty in its mass
ratio estimate \citep[$0.71^{+0.09}_{-0.14}$;][]{me-2397a} due to the
fact that it is composed of a main-sequence star and a field brown
dwarf.  We computed Spearman's rank correlation coefficient in a Monte
Carlo fashion as described above and found $r_S = -0.09\pm0.10$,
showing that there is no evidence for a correlation of eccentricity
with mass ratio over the range of our sample ($0.6 \lesssim q \lesssim
1.0$).  In the absence of such a correlation, any mass ratio selection
effects present in our sample should therefore not induce a systematic
eccentricity bias.

\subsection{Comparison to Solar-type Binaries \label{sec:solar}}

\citet{1991A&A...248..485D} have examined in detail the eccentricity
distribution of a well-defined sample of solar-type binaries. In the
following analysis, we exclude from consideration three of the
binaries in their sample that have an orbit quality grade of 4 or 5
assigned by \citet{1983PUSNO..24g...1W}, which indicates an
indeterminate orbit. Not surprisingly, these binaries (HD~16895AB,
HD~78154AB, and HD~146361AB) are the longest period orbits in the
\citet{1991A&A...248..485D} sample, with orbital periods $>1000$~yr.
(We note that \citet{2010ApJS..190....1R} have recently updated this
volume-limited, solar-type sample, but we cannot use this for
comparison since they did not publish the eccentricity values of their
sample binaries.)

We find that the overall eccentricity distribution of the
\citet{1991A&A...248..485D} solar-type binary sample is very similar
to our orbit sample. To assess this quantitatively, we performed
Kolmogorov-Smirnov (KS) tests in a Monte Carlo fashion. We accounted
for eccentricity errors in our sample by drawing the data in each
trial from the observed distributions (our MCMC chains when possible,
Gaussians otherwise).  We computed the KS statistic ($D$) and the
corresponding probability for every trial. The resulting rms scatter
in $D$ was relatively small($\leq0.05$).\footnote{The definition of
  $D$ is the maximum difference between the two cumulative
  distribution functions that are being tested.} For the final
probability that two samples were drawn from the same parent
distribution, we used the mean of the probabilities from individual
trials. Table~\ref{tbl:ks} shows the resulting probabilities comparing
subsamples of our orbits to subsamples of the solar-type binaries from
\citet{1991A&A...248..485D}.  In general, we find good agreement
between the two samples' eccentricity distributions. However, a more
detailed examination of these samples shows some significant
differences.

Unlike the very low-mass binaries, solar-type binaries follow a strong
correlation with orbital period such that short-period ($<10^3$~days)
binaries span a different range of eccentricities than do long-period
($>10^3$~days) binaries (Figure~\ref{fig:compare}).
\citet{1991A&A...248..485D} suggested that this implies a different
formation mechanism and/or dynamical history for these two period
regimes of solar-type binaries. The same trend has also been reported
for pre--main-sequence, solar-type binaries \citep[][and references
therein]{1994ARA&A..32..465M}, implying that this period--eccentricity
correlation is determined early in a binary's formation history. We
confirm the significance of the correlation in the
\citet{1991A&A...248..485D} sample by computing Spearman's rank
correlation coefficient for solar-type binaries (excluding those that
lie below the critical period for tidal circularization; $\Pcirc =
11.6$~days). The resulting value of $r_S = 0.32$ for this sample of 46
solar-type binaries gives a 98.6\% confidence that $P$ and $e$ are
correlated. As discussed above, our sample shows no such correlation,
which is an indication that the solar and very low-mass populations
are in fact dynamically distinct from one another.

We also note that among the large sample of solar-type binaries from
\citet{1991A&A...248..485D} there is not a single orbit with $e<0.1$
beyond an orbital period of 22~days,\footnote{Solar-type binaries with
  $P>22$~days and $e<0.1$ are indeed known to exist, but none are
  included in the well-defined, volume-limited sample of
  \citet{1991A&A...248..485D}.  For example, in the SB9 catalog
  \citep{2004A&A...424..727P} 20\% of orbits with $P>100$~days and
  quality grades $>3$ have $e<0.1$.  However, such spectroscopic
  binary catalogs represent a compilation of heterogeneous data not
  intended to be used for statistical studies, as they likely harbor
  significant selection biases (e.g., as discussed in
  Section~\ref{sec:sb}).  In addition, we note that the
  \citet{2010ApJS..190....1R} solar-type sample contains 3 such
  binaries according to their Figure~14.} while there are at least 3
such orbits in our sample of 16 binaries (one more,
2MASS~J1534$-$2952AB, could also have $e<0.1$, but the error in its
eccentricity makes this uncertain). This feature of solar-type systems
has also been pointed out for pre--main-sequence binaries \citep[][and
references therein]{1994ARA&A..32..465M}. We used a Fisher's exact
test to compute the probability that this difference in $e<0.1$ orbits
between our sample and the \citet{1991A&A...248..485D} sample arises
solely by chance (i.e., the $P$-value). In order to exclude any
solar-type binaries that may be affected by tidal circularization but
have not reached $e=0$ yet, we consider only binaries with
$P>100$~days (a less conservative, shorter period cutoff could be used
but would not change the results). At these orbital periods, 0 of 39
solar-type binaries have $e<0.1$, while 3 of 14 very low-mass binaries
have eccentricities confidently $<0.1$. Comparing these two samples,
Fisher's exact test finds that this difference is significant with a
$P$-value of 0.0155 (2.2$\sigma$). The $P$-value remains significant
($<0.05$) for any choice of the eccentricity division point ranging
from 0.10 to 0.15. This strong preference for near-circular orbits
among very low-mass binaries provides additional evidence that their
dynamical history may be different from solar-type binaries.

We next consider the location in the $P$--$e$ diagram of very low-mass
binaries that are members of hierarchical triple systems as compared
to their solar-type counterparts. The interpretation given by
\citet{1991A&A...248..485D} and \citet{2010ApJS..190....1R} is that
solar-type binaries that are members of hierarchical triple or
quadruple systems tend to lie near the upper envelope of eccentricity
at a given orbital period. However, examination of the $P$--$e$
diagram of solar-type binaries presented in Figure~14 of
\citet{2010ApJS..190....1R} shows that such members of hierarchical
systems actually occupy a wide range of eccentricities at periods
$>100$~days. In fact, members of hierarchical triple and quadruple
systems are the only near-circular ($e < 0.1$) solar-type binaries at
these periods, suggesting that their higher order nature may be the
cause of their unusually low eccentricities. This is consistent with
the fact that the only such long-period, near-circular binary in the
pre--main-sequence sample of \citet{1994ARA&A..32..465M} is now known
to have a tertiary companion \citep[GW~Ori;][]{2011arXiv1103.3888B}.
In contrast to solar-type binaries, the higher order multiples in our
sample do not populate the upper eccentricity envelope at any period.
In fact, the highest eccentricity orbit in our sample is a true
binary, and the lowest eccentricity orbit belongs to a member of a
triple system. This latter datum may be partially consistent with the
fact that near-circular ($e<0.1$) solar-type binaries tend to be
members of higher order multiples. However, there remains the mystery
as to why such near-circular orbits are much more common among the
true binaries in our sample than in the solar-type binary sample.

Finally, we note that the shortest period binaries in our sample are
not circularized, despite having shorter periods than the empirical
limit from field solar-type binaries ($\Pcirc = 11.6$~days). This was
discussed by \citet{1999AJ....118.2460B}, who used brown dwarf
structural models along with the circularization theory of
\citet{1989A&A...223..112Z} to show that PPl~15 ($P = 5.8$~days) is
not expected to ever tidally circularize as its critical period is
$\sim$4.5~days. By extension, the orbit of the eclipsing binary
2MASS~J0535$-$0546 ($P = 9.8$~days), which has even lower masses and
thus a shorter critical period, is also predicted to never
circularize, so its nonzero eccentricity is not merely a symptom of
its youth \citep[$\sim$1~Myr;][]{2006Natur.440..311S}.

\subsection{Comparison to Theory \label{sec:theory}}

\citet{ambartsumian37} showed that if binary companions are
distributed in phase space solely as a function of energy, whether
according to the Boltzmann distribution or any other arbitrary
function, the cumulative distribution of eccentricities must be
proportional to $e^2$ and the mean value of $e$ should be equal to
$\frac{2}{3}$. He argued that the observational data for wide stellar
binaries at the time was in strong disagreement with this prediction,
implying that binaries neither originate from nor relax via dynamical
interactions into a state of statistical equilibrium.
\citet{1991A&A...248..485D} tested the \citet{ambartsumian37}
distribution with modern data for solar-type binaries, applying a
correction due to eccentricity bias from \citet{1977PASP...89..400H},
and claimed that the $e^2$ distribution described the wider binaries
($P > 10^3$~days) reasonably well. However, their data still possess a
severe lack of high-eccentricity binaries, especially after excluding
the three binaries with $>$1000-yr orbits, as these happen to include
two of the most eccentric orbits in their sample ($e\gtrsim0.8$).

Our sample seems to also contravene the $e^2$ distribution, as
high-$e$ binaries are notably scarce. To quantify this, we performed
KS tests as described in Section~\ref{sec:solar}. Following the
discussion in Section~\ref{sec:vb}, we corrected for visual binary
orbit bias by adding simulated data points uniformly distributed in
each of the last two bins ($e$ = 0.6--0.8 and 0.8--1.0). These bins
contain 0 and 1 visual binary, respectively, and after accounting for
the bias the number in each bin was 0 and 2.4 binaries, averaged over
$10^4$ trials. The resulting probability that the null hypothesis is
correct (i.e., the observed sample is drawn from the predicted $e^2$
distribution) was $9.1\times10^{-5}$ for the entire sample and
$2.2\times10^{-3}$ when considering only the visual binaries in our
sample. Because of the potential impact of our assumed bias correction
on the results, we also performed KS tests considering only
eccentricities below 0.6, for which there is expected to be negligible
observational bias. These tests gave probabilities of
$5.3\times10^{-7}$ and $7.1\times10^{-5}$ for the full sample and
visual binaries, respectively. Thus, we conclude that our sample of
very low-mass binaries does not follow an \citet{ambartsumian37}
distribution over orbital periods of 5.8 to $8\times10^3$~days.

\begin{figure}

\centerline{
\includegraphics[width=3.2in,angle=0]{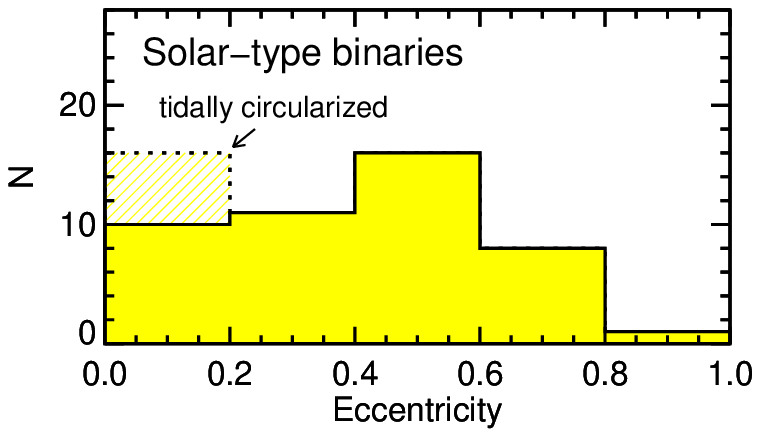}
\hskip -0.1in
\includegraphics[width=3.2in,angle=0]{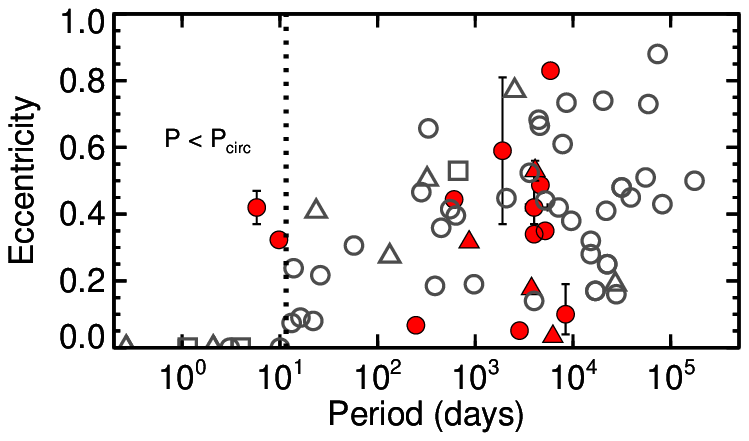}}
\vskip 0.1in
\centerline{
\includegraphics[width=3.2in,angle=0]{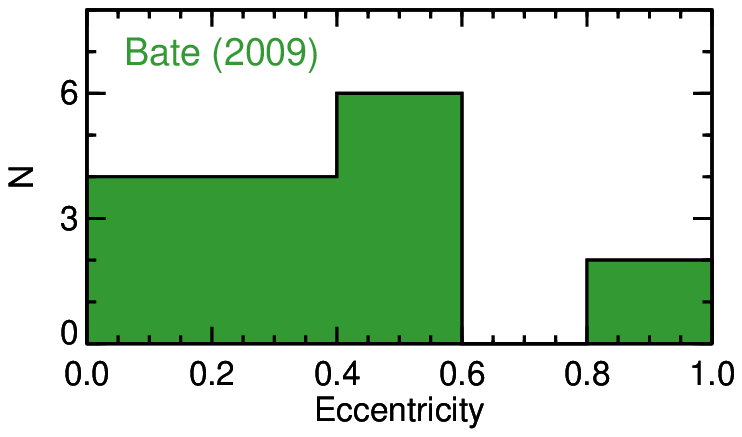}
\hskip -0.1in
\includegraphics[width=3.2in,angle=0]{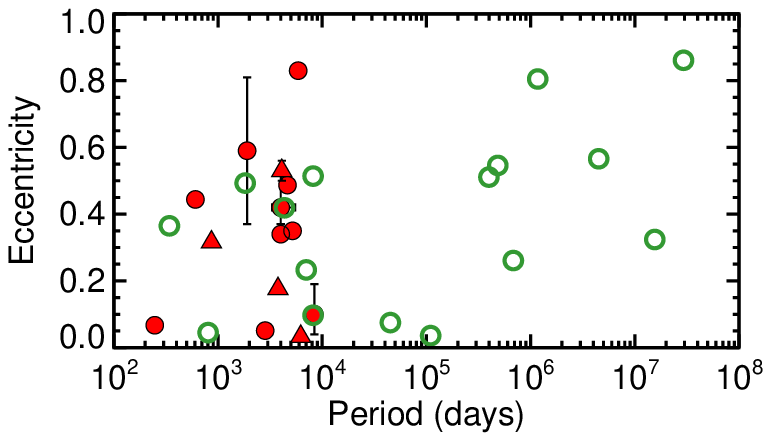}}
\vskip 0.1in
\centerline{
\includegraphics[width=3.2in,angle=0]{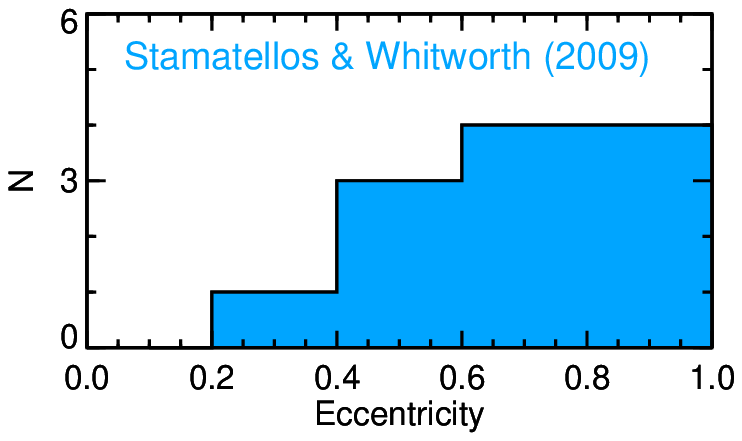}
\hskip -0.1in
\includegraphics[width=3.2in,angle=0]{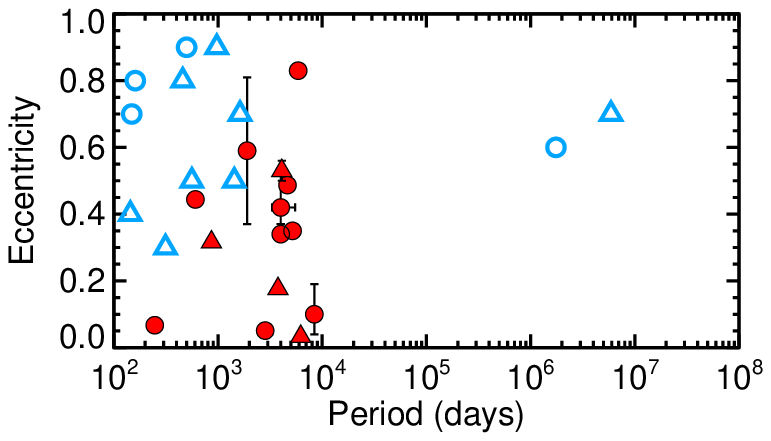}}

\caption{ Eccentricity distributions (left) and orbital
  period--eccentricity diagrams (right, open symbols) for various
  samples.  \emph{Top:} Solar-type binaries from
  \citet{1991A&A...248..485D}, with three low-quality, $>$1000-yr
  orbits excluded. Tidally circularized orbits ($P < \Pcirc$) are
  plotted separately in the histogram (hatched region).  Triangles and
  squares denote members of hierarchical triple and quadruple systems,
  respectively.  \emph{Middle:} Cluster formation model simulated
  binaries from \citet{2009MNRAS.392..590B}; eccentricities for this
  sample of very low-mass simulated binaries are reported by
  \citet{2010HiA....15..769B}.  \emph{Bottom:} Circumstellar disk
  fragmentation model simulated binaries from
  \citet{2009MNRAS.392..413S}, with triangles denoting binaries that
  remain bound to the simulated host star.  The very low-mass binaries
  in our sample are shown as filled red symbols on all $P$--$e$
  diagrams for comparison. \label{fig:compare}}

\end{figure}

We have also compared our sample to the predictions of numerical
simulations that strive to describe the formation of such low-mass
binaries. \citet{2009MNRAS.392..590B} simulated the collapse of a
500~\Msun\ molecular cloud, producing sufficiently large numbers of
stars and brown dwarfs ($>10^3$) to do statistical comparisons with
observations. \citet{2010HiA....15..769B} reported the parameters of
sixteen binaries with total masses of $<0.2$~\Msun\ that were
generated in this simulation, and we plot their orbital periods and
eccentricities in Figure~\ref{fig:compare}.\footnote{Note that these
  are results from the ``rerun'' simulation, which had a smaller sink
  radius of 0.5-AU, no gravitational softening, and a shorter run time
  than the main calculation in \citet{2009MNRAS.392..590B}.} Our
observed eccentricity distribution agrees very well with the
simulation of \citet{2009MNRAS.392..590B}, with a 83\% probability of
both samples being drawn from the same parent distribution according
to a KS test (92\% for visual binaries alone), and this result is not
changed if we consider only those simulated binaries within our
sample's semimajor axis upper limit of 3.7~AU. However, we note that
\citet{2009MNRAS.392..590B} does not produce any high-$e$ ($>0.6$)
binaries over the same orbital period range as we observe, as all of
his high-$e$ binaries have very long orbital periods ($>10^6$~days).
This could be due to the small number statistics of the simulation
results, but if not then it is suggestive of a disagreement with the
observations that is only made stronger by the fact that the orbit
bias correction implies a total of $\approx$3 binaries with $e>0.6$
for our sample. This discrepancy could be due to the fact that at the
end of the simulation the \citet{2009MNRAS.392..590B} binaries are
still evolving toward shorter orbital periods by dynamical
interactions with other protostars and their own accretion disks.
Thus, the high-$e$ systems could ultimately reach the orbital periods
of our sample. However, their eccentricities would also likely damp
down in the process, and thus the discrepancy with observations would
remain. Another significant concern is that the very low-mass stars
and brown dwarfs in the simulation of \citet{2009MNRAS.392..590B} are
formed largely by being ejected from their initial gas reservoirs
through dynamical interactions.  \citet{2009MNRAS.392.1363B} showed
that by including a more realistic equation of state and radiative
feedback, such ejection events become much less common, which resolves
the problem that previous simulations (including
\citealp{2009MNRAS.392..590B}) produced far too many brown dwarfs. It
remains to be seen how this improvement in the input physics will
impact the orbital elements of binaries produced in these simulations.

In recent simulations by \citet{2009MNRAS.392..413S}, very low-mass
stars, brown dwarfs, and planetary-mass objects are formed via
gravitational instability in the outer parts ($\gtrsim 100$~AU) of
massive circumstellar disks ($M_{\rm disk}/M_{\star} = 1.0$). The
orbital periods and eccentricities of the twelve binaries with $\Mtot
< 0.2$ that were generated in these simulations are shown in
Figure~\ref{fig:compare}. These binaries are a mix of three field
binaries (those that are ejected from the disk), and nine binaries
that remain bound to the star in a hierarchical triple. The
eccentricities of these two subsets are not statistically different
from each other as shown by a KS test ($D = 0.5$, 38\% chance of being
drawn from the same parent distribution), and so we simply compare the
full set of simulated binaries to our sample. The simulated binaries
have significantly higher eccentricities than we observe, with our KS
test giving a probability of only $9.9\times10^{-3}$ that the two
samples were drawn from the same parent distribution. This result is
not changed significantly if we only consider simulated binaries
within our sample's semimajor axis upper limit of 3.7~AU. As discussed
by \citet{2009MNRAS.392..413S}, the binary components formed in these
simulations have their own disks, and after the initial 20~kyr their
simulations ignore the gas in order to follow the long term evolution
of the system.  The tidal interactions associated with these disks
that are not accounted for in the simulations at late times could damp
the eccentricities of the simulated binaries.  Indeed,
\citet{2009MNRAS.392..590B} found that including such dissipative
interactions was critical in reproducing the stellar eccentricity
distribution. Thus, including such additional input physics may solve
the disagreement between our observed eccentricities and the
simulations of \citet{2009MNRAS.392..413S}. On the other hand, such
large eccentricities may simply be a hallmark of forming in the
frenetic dynamical environment of a fragmenting circumstellar disk.


\section{Conclusions}

We have considered for the first time the eccentricity distribution of
a large sample of orbits for very low-mass star and brown dwarf
binaries. We have examined potential observational biases in this
sample, and for visual binaries we show through Monte Carlo
simulations that by choosing appropriate selection criteria all
eccentricities are equally represented ($\lesssim5\%$ difference
between input and output eccentricity distributions). We find that our
sample of very low-mass binaries populate a broad range in
eccentricity from nearly circular to highly eccentric ($0.03 < e <
0.83$), showing a strong preference for modest eccentricities (the
mean and median of the full sample is $e=0.34$).  The binaries among
our sample that are members of hierarchical triple systems show no
special preference for their eccentricity as compared to the other
binaries, consistent with a common formation mechanism for both
populations. The overall eccentricity distribution of our sample is
similar to solar-type binaries from \citet{1991A&A...248..485D}.
However, very low-mass binaries lack any correlation between
eccentricity and period or semimajor axis, which is one of the key
features of the solar-type population. In addition, near-circular
orbits appear to be quite common among very low-mass binaries, as at
least 3 of our 16 sample binaries have $e<0.1$.  However, such orbits
are apparently rare among solar-type binaries (in the
\citealp{1991A&A...248..485D} sample, none of the 42 binaries beyond
an orbital period of 22~days have $e<0.1$).  This difference between
the two samples is statistically significant according to Fisher's
exact test ($P$-value of 0.0155). Thus, the orbital data currently
available are suggestive of a different formation mechanism or
dynamical history for very low-mass binaries as compared to solar-type
stars.

We have also tested formation models by comparing our sample of very
low-mass binaries to those produced in numerical simulations. The
cluster formation model of \citet{2009MNRAS.392..590B} predicts an
overall eccentricity distribution that is well-matched to the data.
However, there is a potential discrepancy with the observations due to
the lack of any eccentric binaries ($e>0.6$) at orbital periods
corresponding to our sample. This may be caused by the same physical
mechanism that is needed to bring binaries formed in the simulation to
these orbital periods, namely tidal damping by the accompanying gas
disks. In contrast, the gravitational instability model of
\citet{2009MNRAS.392..413S} produces too many high-$e$ binaries, with
an absence of \emph{any} modest eccentricity ($e<0.3$) orbits. This
may be due to the fact that tidal dissipation by gas disks is not
included in the \citet{2009MNRAS.392..413S} simulation, causing the
resulting eccentricity distribution to be inconsistent with our sample
at high significance. A third possibility is that the observed
discrepancies are simply due to multiple formation processes being
responsible for producing the field population of very low-mass
binaries, so no one model matches all the data. A further complication
is that binaries formed in a cluster environment are likely to undergo
additional dynamical processing after their initial eccentricities are
set, and thus $N$-body simulations are needed to assess the impact of
such effects on the final eccentricity distribution.

To improve on our observational tests, future efforts should aim to
cover a wider range of orbital parameter space, particularly expanding
the sample at short orbital periods ($P\lesssim10^3$~days). The
discovery of such tight binaries is only currently feasible from RV
monitoring or the ``spectral binary'' technique (i.e., identifying the
signature of a fainter companion of disparate spectral type in the
integrated light spectrum). For example, \citet{2008ApJ...681..579B}
used this technique to identify 2MASS~J0320$-$0446 as a binary
independent of its identification as an SB1
\citep{2008ApJ...678L.125B}. However, even for such spectral binaries
the determination of orbits requires either precise RV measurements or
resolved astrometry. This will require new capabilities (precision
near-infrared radial velocities for L and T dwarfs) as well as new
facilities (e.g., TMT/GMT/E-ELT and \JWST\ to resolve the much smaller
semimajor axes). Fortunately, future binaries that will fill in the
bottom two orders of magnitude in orbital period will not require the
same duration of patient orbital monitoring that has been necessary
over the last decade to develop the current sample of $\sim$3--30-yr
period visual orbits.

We have taken the first look at what clues the ensemble of orbital
eccentricities of very low-mass stars and brown dwarfs reveal about
their formation.  This work follows on previous statistical studies of
the visual binary population that found distributions of mass ratio
and semimajor axis that were significantly different than for
solar-type binaries \citep[e.g.,][]{2003ApJ...587..407C,
  2007prpl.conf..427B}.  The theoretical implication that disks may be
critical in shaping the eccentricity distribution is in accord with
the fact that eccentricity is closely associated with a binary's
angular momentum. This potential relationship is intriguing, as it
suggests the possibility for disk-related observational tests. For
example, if high-eccentricity binaries are simply those that lost
their disks early in their formation, then perhaps the binary
components rotate more rapidly than lower eccentricity binary
components due to a lack of disk-braking.  Observational evidence has
also shown that the formation environment may play an important role
in determining the orbits of binaries, with dynamical interactions in
the dense Orion Nebula Cluster likely responsible for sculpting the
semimajor axis distribution of stellar binaries
\citep{2007AJ....134.2272R} while binaries in the sparser Taurus
region seem to be able to survive at much wider separations
\citep{2009ApJ...703.1511K}. Further theoretical modeling of the
binary formation process taking such effects into account will enable
us to better understand the physical processes influencing observed
orbital eccentricities.


\acknowledgments

Our research has employed NASA's Astrophysical Data System; and the
SIMBAD database operated at CDS, Strasbourg, France.  We are grateful
to Matthew Bate for providing us with the orbital elements of the
lowest mass binaries in his simulation.  We have benefitted from
discussions with Adam Kraus about binary formation, John Johnson about
RV selection effects, and Tsevi Mazeh about orbital dynamics.  TJD and
MCL acknowledge support for this work from NSF grants AST-0507833 and
AST-0909222. TJD acknowledges support from Hubble Fellowship grant
HST-HF-51271.01-A awarded by the Space Telescope Science Institute,
which is operated by AURA for NASA, under contract NAS 5-26555.
Finally, the authors wish to recognize and acknowledge the very
significant cultural role and reverence that the summit of Mauna Kea
has always had within the indigenous Hawaiian community.  We are most
fortunate to have the opportunity to conduct observations from this
mountain.

{\it Facilities:} \facility{Keck II Telescope (LGS AO, NGS AO, NIRC2)}


\appendix

\section{Converting Projected Separation to Semimajor Axis \label{app:sma-corr}}

A general problem in the study of visual binary stars is that orbit
determinations are often infeasible, so the true semimajor axis ($a$)
of the binary must be estimated from the projected separation on the
sky ($\rho$).  In the usual approach, $\rho$ is multiplied by a
correction factor that accounts for the fact that the observed
separation will typically be smaller than the true semimajor axis.
This is always true for circular orbits, but eccentric orbits can be
observed at much wider separations than the actual semimajor axis.
\citet{1976ApJS...30..273A} used a correction factor of $a/\rho =
\frac{\pi}{2}$, computed analytically by assuming edge-on circular
orbits.  \citet{1992ApJ...396..178F} performed a Monte Carlo
simulation that assumed random inclinations and reported 1.26 as their
time-averaged correction factor.  Unfortunately, they did not describe
the input eccentricity distribution used nor did they report
confidence limits for the correction factor.
\citet{1999PASP..111..169T} performed several Monte Carlo simulations
assuming random viewing angles and a variety input eccentricity
distributions.  He showed that the conversion factor not only varies
greatly for different eccentricities, but that the probability
distributions for these values are quite broad and can display one or
more peaks.  Thus, it is important to account for these large
uncertainties when converting $\rho$ to $a$.

\begin{figure}

\centerline{\includegraphics[width=6.0in,angle=0]{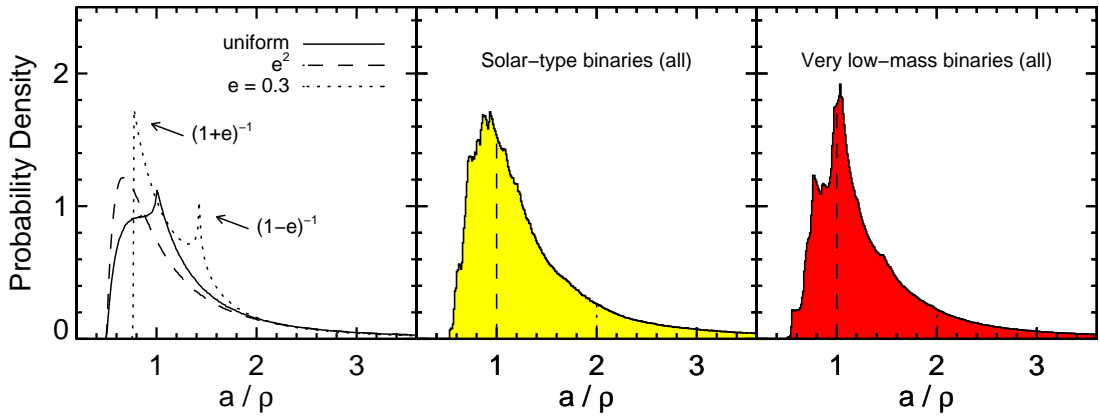}}

\caption{Distribution of conversion factors from projected separation
  ($\rho$) to true semimajor axis ($a$) for a variety of eccentricity
  distributions (see Table~\ref{tbl:sma-corr}).  The left panel shows
  the distribution at an arbitrary fixed eccentricity ($e = 0.3$), as
  well as for two analytical eccentricity distributions: uniform and
  $e^2$ \citep{ambartsumian37}. For the solar-type binaries (middle)
  and very low-mass binaries (right), we show only the probability
  distributions for the full samples because subsets of these samples
  give essentially the same results.  The dotted and dashed lines in
  the right two panels show where the distributions are truncated in
  the cases of moderate ($a/\rho = 2$) and severe ($a/\rho = 1$)
  discovery bias, respectively. \label{fig:sma-corr}}

\end{figure}

With empirical eccentricity distributions in hand for solar-type and
very low-mass binaries, we have undertaken Monte Carlo simulations to
compute appropriate conversion factors for binaries drawn from these
populations.  We assumed uniformly distributed observation times ($0 <
t-t_0 < P$), uniform arguments of periastron ($0\degree < \omega <
360\degree$), and randomly oriented orbits ($p(i) \propto \sin{i}$).
From $10^7$ Monte Carlo trials we computed observed projected
separations ($\rho$) and thus arrived at a distribution of conversion
factors ($a/\rho$).  The results for various assumptions about the
input eccentricity distribution are given in Table~\ref{tbl:sma-corr}.
In Figure~\ref{fig:sma-corr} we show the probability distributions of
these conversion factors.  For comparison, we also show the resulting
distribution if $e$ is held constant in the simulations.  In this
case, the distribution has two peaks: one at $1/(1+e)$ corresponding
to its apastron passage and one at $1/(1-e)$ corresponding to
periastron.  The empirical distributions of $a/\rho$ have many small
localized peaks due to the fact that they are composed of many
individual eccentricity measurements that each contribute two peaks.
The effect is more pronounced for the smaller sample of the thirteen
very low-mass binaries as compared to the 46 solar-type binaries.

We have further considered the impact of discovery bias (see
Section~\ref{sec:vb}) on these distributions. The resolution limit of
an imaging survey will preclude the discovery of binaries when they
are at very small separations, which results in truncating the tail of
conversion factors at large $a/\rho$. In Table~\ref{tbl:sma-corr} we
give the results when considering two cases of discovery bias. In the
moderate case, the inner working angle (IWA) is a factor of two
smaller than the true semimajor axis (IWA = $a/2$), which truncates
the conversion factor distribution at $a/\rho = 2$. In the severe case
the resolution limit is equal to the semimajor axis, so only eccentric
binaries can be discovered, and the conversion factors are never
greater than $a/\rho = 1$. These truncation values are shown as
vertical lines in Figure~\ref{fig:sma-corr}.



\clearpage
\begin{deluxetable}{lcccc}
\tablecaption{New Keck AO Astrometry \label{tbl:astrom}}
\tablewidth{0pt}
\tablehead{
\colhead{Epoch (UT)} &
\colhead{$\rho$ (mas)} &
\colhead{PA (\degree)} &
\colhead{Filter} &
\colhead{$\Delta{m}$ (mag)}}
\startdata

\multicolumn{5}{c}{\bf LP~415-20AB (LGS)} \\
\cline{1-5}
 2008 Jan    15 &   $89.3\pm0.4$      &   $56.0\pm0.5$ \phn     & \Ks &  $0.60\pm0.07$  \\
 2008 Sep \phn8 &  $107.5\pm0.8$\phn  &   $61.0\pm0.3$ \phn     & \Ks & $0.589\pm0.013$ \\
 2009 Sep    28 & $129.81\pm0.15$\phn &  $66.70\pm0.06$\phn     & \Ks & $0.524\pm0.022$ \\
 2010 Jan \phn9 & $134.96\pm0.19$\phn &  $68.05\pm0.05$\phn     & \Ks & $0.548\pm0.016$ \\
\cline{1-5}
\multicolumn{5}{c}{} \\
\multicolumn{5}{c}{\bf HD~130948BC (NGS)} \\
\cline{1-5}
 2010 Jan \phn9 &   $89.0\pm0.4$      & $318.21\pm0.25$\phn\phn & \Ks & $0.212\pm0.023$ \\
 2010 Mar    22 & $100.15\pm0.15$\phn & $316.95\pm0.07$\phn\phn & $K$ & $0.206\pm0.007$ \\

\enddata

\end{deluxetable}

\clearpage
\begin{deluxetable}{lcc}
\tablecaption{Updated Orbit Determinations \label{tbl:orbits}}
\tablewidth{0pt}
\tablehead{
\colhead{Parameter} &
\colhead{LP~415-20AB} &
\colhead{HD~130948BC}}
\startdata

$P$ (yr)                       &   $14.4\pm0.4$      \phn&  $10.28\pm0.17$        \phn \\
$a$ (mas)                      &$93.5^{+3.1}_{-2.3}$ \phn&   $124.5\pm1.4$    \phn\phn \\
$e$                            &$0.708^{+0.016}_{-0.014}$\phn&   $0.176\pm0.006$           \\
$i$ (\degree)                  &  $59.9\pm2.2$       \phn&    $95.7\pm0.2$        \phn \\
$\Omega$ (\degree)             &$261.5^{+2.4}_{-2.7}$\phn\phn&  $313.39\pm0.12$   \phn\phn \\
$\omega$ (\degree)             &   $351\pm7$     \phn\phn&     $244\pm3$          \phn \\
$T_0$ (MJD)                    & $53505\pm80$\phn\phn\phn&  $ 54613\pm22$ \phn\phn\phn \\
\Mtot\tablenotemark{a} (\Msun) &     \nodata             &  $0.1095\pm0.0022$          \\
$\chi^2$ (DOF)                 &    6.83  (7)            &       9.38 (13)             \\

\enddata

\tablenotetext{a}{The total mass accounts for the uncertainty in the
  parallax.  LP~415-20AB has no measured parallax and thus no direct
  mass measurement.}

\end{deluxetable}

\clearpage
\begin{deluxetable}{lccccccccccccc}
\tabletypesize{\scriptsize}
\rotate
\tablecaption{Visual Binaries with Published Orbits \label{tbl:vb}}
\tablewidth{0pt}
\tablehead{
\colhead{Name} &
\colhead{Sp.\ Type} &
\multicolumn{2}{c}{Distance} &
\colhead{} &
\multicolumn{3}{c}{Orbit} &
\colhead{} &
\multicolumn{2}{c}{Orbit Quality} &
\colhead{} &
\colhead{IWA} &
\colhead{IWA/$a$} \\
\cline{3-4}
\cline{6-8}
\cline{10-11}
\colhead{} &
\colhead{Primary} &
\colhead{$d$ (pc)} &
\colhead{Ref.} &
\colhead{} &
\colhead{$a$ (AU)} &
\colhead{$e$} &
\colhead{Ref.} &
\colhead{} &
\colhead{$\Delta t_{\rm obs}/P$} &
\colhead{$\sigma_M/M$} &
\colhead{} &
\colhead{(\arcsec)} &
\colhead{}}
\startdata

Gl~569Bab            & M8.5 &\phn  $9.65\pm0.16$            & 16 & &     $0.923\pm0.018$     &      $0.316\pm0.005$      &   8  & &    334\%    &   $-$3\%, +4\%   & &  0$\farcs$05  &  0.52  \\
2MASS~J0920$+$3517AB & T0   &  $25.4^{+0.8}_{-0.7}$         &  9 & &  $1.67^{+0.06}_{-0.05}$ &      $0.217\pm0.019$      &   9  & &    144\%    &        3\%       & &  0$\farcs$06  &  0.91  \\
LP~349-25AB          & M8   &  $13.2\pm0.3$                 & 10 & &      $1.94\pm0.04$      &      $0.051\pm0.003$      &   8  & &     76\%    &      0.6\%       & &  0$\farcs$10  &  0.68  \\
DENIS~J2252$-$1730AB & L6   &  $15.0\pm0.3$                 &  9 & &   $2.2^{+0.5}_{-0.2}$   &   $0.42^{+0.11}_{-0.08}$  &   9  & &  34\%--56\% &        5\%       & &  0$\farcs$07  &  0.48  \\
HD~130948BC          & L4   & $18.17\pm0.11$                & 16 & &      $2.26\pm0.03$      &      $0.176\pm0.006$      &   1  & &     73\%    &       0.8\%      & &  0$\farcs$06  &  0.48  \\
2MASS~J2132$+$1341AB & L5   &  $27.5\pm0.5$                 &  9 & &      $2.24\pm0.10$      &   $0.34^{+0.03}_{-0.02}$  &   9  & &     35\%    &        5\%       & &  0$\farcs$05  &  0.61  \\
$\epsilon$~Indi~Bab  & T1   &\phn $3.624\pm0.004$           & 16 & &      $2.43\pm0.11$      &       $0.53\pm0.03$       &   2  & &     38\%    &       0.9\%      & &  0$\farcs$08  &  0.12  \\
LP~415-20AB          & M7.5 &$[27\pm4]$\tablenotemark{a}    &  1 & &      $[2.5\pm0.4]$      & $0.708^{+0.016}_{-0.014}$ &   1  & &     55\%    &   $-$7\%, +9\%   & &  0$\farcs$12  &  1.28  \\
2MASS~J0746$+$2000AB & L0   & $12.21\pm0.04$                &  5 & &     $2.899\pm0.013$     &      $0.487\pm0.004$      & 1,11 & &     68\%    &        2\%       & &  0$\farcs$06  &  0.25  \\
LHS~2397aAB          & M8   &  $14.3\pm0.4$                 & 13 & &      $3.09\pm0.10$      &      $0.350\pm0.005$      &   7  & &     83\%    &        3\%       & &  0$\farcs$12  &  0.56  \\
2MASS~J1534$-$2952AB & T5   & $13.59\pm0.22$                & 15 & &   $3.2^{+0.4}_{-0.3}$   &   $0.10^{+0.09}_{-0.06}$  & 1,11 & &  32\%--44\% &        2\%       & &  0$\farcs$06  &  0.26  \\
LHS~1070BC           & M8.5 &\phn  $7.72\pm0.15$            &  3 & &      $3.56\pm0.07$      & $0.033^{+0.003}_{-0.002}$ & 1,14 & &     82\%    &       0.9\%      & &  0$\farcs$10  &  0.22  \\
LHS~1901AB           & M6.5 &  $12.9\pm0.5$                 & 12 & &      $3.70\pm0.16$      & $0.194^{+0.025}_{-0.021}$ &   8  & &     39\%    &       1.4\%      & &  0$\farcs$10  &  0.35  \\
2MASS~J2140$+$1625AB & M8.5 &$[26\pm5]$\tablenotemark{a}    &  1 & &      $[3.7\pm0.7]$      &   $0.36^{+0.14}_{-0.10}$  & 1,11 & &  38\%--57\% &  $-$27\%, +53\%  & &  0$\farcs$12  &  0.83  \\
2MASS~J0850$+$1057AB & L7   &     $38^{+7}_{-5}$            & 17 & &   $4.8^{+3.9}_{-1.4}$   &       $0.64\pm0.26$       &  11  & &  10\%--49\% &       100\%      & &  0$\farcs$06  &  0.48  \\
2MASS~J1728$+$3948AB & L5   &   $24.1^{+2.1}_{-1.8}$        & 17 & &       $5.3\pm0.8$       &   $0.28^{+0.35}_{-0.28}$  &  11  & &  20\%--47\% &  $-$27\%, +167\% & &  0$\farcs$06  &  0.27  \\
2MASS~J2206$-$2047AB & M8   &  $26.7^{+2.6}_{-2.1}$         &  4 & &   $5.8^{+0.8}_{-0.7}$   &       $0.25\pm0.08$       &  14  & &  20\%--28\% &       2.0\%      & &  0$\farcs$12  &  0.56  \\
2MASS~J1847$+$5522AB & M6.5 &$[25\pm5]$\tablenotemark{a}    &  1 & &      $[5.8\pm1.5]$      &    $0.1^{+0.5}_{-0.1}$    &  11  & &   9\%--23\% &  $-$72\%, +194\% & &  0$\farcs$08  &  0.34  \\
2MASS~J1750$+$4424AB & M7.5 &$[29\pm8]$\tablenotemark{a}    &  1 & &       $[21\pm13]$       &       $0.71\pm0.18$       &  11  & &  1.3\%--9\% &       60\%       & &  0$\farcs$12  &  0.16  \\
2MASS~J1426$+$1557AB & M8.5 &$[25\pm5]$\tablenotemark{a}    &  1 & &       $[60\pm40]$       &   $0.85^{+0.10}_{-0.41}$  &  11  & & 0.2\%--20\% &  $-$100\%, +73\% & &  0$\farcs$12  &  0.05  \\

\enddata

\tablecomments{This table includes all resolved binaries with a total
  mass $<0.2$~\Msun\ for which an orbit determination has been
  published. Orbit quality metrics are given for each binary:
  (1)~fraction of orbital phase observed ($\Delta t_{\rm obs}/P$), if
  the period uncertainty is $>10\%$ the 1$\sigma$ range is given; and
  (2)~fractional 1$\sigma$ error in the total mass from the orbit fit
  alone ($\sigma_M/M$), i.e., independent of the uncertainty in the
  distance.  High quality orbits have a large fraction of the orbit
  covered and/or a low mass error.}

\tablenotetext{a}{Spectrophotometric distance estimate derived
  following the method used in the Appendix of \citet{liu-1209}.}

\tablerefs{(1)~This work; (2)~\citet{2009AIPC.1094..509C};
  (3)~\citet{2005AJ....130..337C}; (4)~\citet{2006AJ....132.1234C};
  (5)~\citet{2002AJ....124.1170D}; (6)~\citet{me-2206};
  (7)~\citet{me-2397a}; (8)~\citet{me-latem}; (9)~\citet{me-phd};
  (11)~\citet{qk10}; (10)~\citet{2009AJ....137..402G};
  (12)~\citet{2009AJ....137.4109L}; (13)~\citet{1992AJ....103..638M};
  (14)~\citet{2008A&A...484..429S}; (15)~\citet{2003AJ....126..975T}.
  (16)~\citet{2007hnrr.book.....V}; (17)~\citet{2004AJ....127.2948V}.}

\end{deluxetable}

\clearpage
\begin{deluxetable}{lccccccc}
\tabletypesize{\scriptsize}
\rotate
\tablecaption{Very Low-Mass Binary Orbit Sample \label{tbl:ecc}}
\tablewidth{0pt}
\tablehead{
\colhead{Name} &
\colhead{Int.-Light} &
\colhead{Binary} &
\colhead{$P$} &
\colhead{$e$} &
\colhead{\Mtot} &
\colhead{Mass ratio} &
\colhead{Orbit} \\
\colhead{} &
\colhead{Sp.\ Type} &
\colhead{Type} &
\colhead{(days)} &
\colhead{} &
\colhead{(\Msun)} &
\colhead{($q \equiv M_2/M_1$)} &
\colhead{Ref.}}
\startdata

PPl~15               &   M6.5   & S2                 &     $5.8\pm0.3$             &  $0.42\pm0.05$            & $[0.124\pm0.004]$\tablenotemark{a} &       $[0.85\pm0.05]$     &  2  \\
2MASS~J0535$-$0546   &   M6.5   & E                  & $9.77962\pm0.00004$         & $0.323\pm0.006$           &  $0.088\pm0.005$                   &        $0.64\pm0.04$      & 12  \\
2MASS~J0320$-$0446   &   M8     & S1                 &   $246.9\pm0.5$     \phn\phn& $0.067\pm0.015$           &  $[0.15\pm0.03]$\tablenotemark{a}  &        [0.60--0.87]       &  3  \\
LSR~J1610$-$0040     &  d/sdM7  & A,S1               &     $607\pm4$       \phn\phn& $0.444\pm0.017$           & $[0.162\pm0.015]$\tablenotemark{a} &        [0.61--0.86]       &  5  \\
Gl~569Bab            &   M8.5   & V\tablenotemark{b} &   $864.5\pm1.1$     \phn\phn& $0.316\pm0.005$           &  $0.140^{+0.009}_{-0.008}$         &$[0.866^{+0.019}_{-0.014}]$&  7  \\
Cha~H$\alpha$~8      &   M5.75  & S1                 &    $1900\pm130$         \phn&  $0.59\pm0.22$            &  $[0.15\pm0.02]$\tablenotemark{a}  &         [0.3--0.7]        &  9  \\
LP~349-25AB          &   M8     & V                  &    $2834\pm15$      \phn\phn& $0.051\pm0.003$           &  $0.120^{+0.008}_{-0.007}$         &$[0.872^{+0.014}_{-0.018}]$&  7  \\
HD~130948BC          &   L4     & V\tablenotemark{b} &    $3760\pm60$      \phn\phn& $0.176\pm0.006$           & $0.1095\pm0.0022$                  &      $[0.948\pm0.005]$    &  1  \\
$\epsilon$~Indi~Bab  &   T2.5   & V\tablenotemark{b} &    $4090\pm180$         \phn&  $0.53\pm0.03$            &  $0.116\pm0.001$                   &       $[0.63\pm0.02]$     &  4  \\
2MASS~J2132$+$1341AB &   L6     & V                  & $4000^{+300}_{-200}$    \phn&  $0.34^{+0.03}_{-0.02}$   &  $[0.13\pm0.02]$\tablenotemark{a}  &       $[0.76\pm0.03]$     &  8  \\
DENIS~J2252$-$1730AB &   L7.5   & V                  & $4000^{+1500\phn}_{-700}$   &  $0.42^{+0.11}_{-0.08}$   &  $[0.12\pm0.02]$\tablenotemark{a}  &       $[0.70\pm0.03]$     &  8  \\
2MASS~J0746$+$2000AB &   L0.5   & V                  &    $4640\pm40$      \phn\phn& $0.487\pm0.004$           &  $0.151\pm0.003$                   &       $[0.94\pm0.02]$     & 1,10\\
LHS~2397aAB          &   M8     & V                  &    $5190\pm40$      \phn\phn& $0.350\pm0.005$           &  $0.146\pm0.014$                   & $[0.71^{+0.09}_{-0.14}]$  &  6  \\
LHS~1901AB           &   M7     & V                  &    $5880\pm180$         \phn& $0.830\pm0.005$           &  $0.194^{+0.025}_{-0.021}$         &$[0.958^{+0.015}_{-0.014}]$&  7  \\
LHS~1070BC           &   M8.5   & V\tablenotemark{b} &    $6220\pm16$      \phn\phn& $0.033^{+0.003}_{-0.002}$ &  $0.156\pm0.009$                   &       $[0.84\pm0.05]$     & 1,11\\
2MASS~J1534$-$2952AB &   T5     & V                  & $8400^{+1500\phn}_{-1100}$  &  $0.10^{+0.09}_{-0.06}$   &  $0.061\pm0.003$                   &      $[0.936\pm0.012]$    & 1,10\\

\enddata

\tablecomments{This table includes visual binaries from
  Table~\ref{tbl:vb} that pass our eccentricity de-biasing criteria
  ($0.9~{\rm AU}<a<3.7$~AU and IWA/$a<0.75$), as well as all
  unresolved binaries with a total mass $<0.2$~\Msun\ for which an
  orbit determination has been published. The type of each binary is
  given: astrometric (A), spectroscopic single-lined (S1) or
  double-lined (S2), eclipsing (E), or visual (V).  All table values
  enclosed in brackets ``[]'' are estimates based on models.  Mass
  ratio estimates for visual binaries are based on Lyon evolutionary
  models \citep{2000ApJ...542..464C, 2003A&A...402..701B}, while the
  estimates for the SB1s and SB2 are taken from the literature
  \citep{1999AJ....118.2460B, 2008ApJ...686..548D,
    2009AJ....137.4621B, 2010A&A...521A..24J}.}

\tablenotetext{a}{No direct mass measurement is available, so
  estimates from the literature based on evolutionary models and/or
  ancillary data are given \citep{1999AJ....118.2460B,
    2003ApJ...598.1265S, 2006AJ....132..891R, 2007AJ....133.2320S,
    2008ApJ...686..548D, 2009AJ....137.4621B, 2010A&A...521A..24J}.}

\tablenotetext{b}{Binary is the short-period component in a
  hierarchical triple system.}

\tablerefs{(1)~This work; (2)~\citet{1999AJ....118.2460B};
  (3)~\citet{2010ApJ...723..684B}; (4)~\citet{2009AIPC.1094..509C};
  (5)~\citet{2008ApJ...686..548D}; (6)~\citet{me-2397a};
  (7)~\citet{me-latem}; (8)~\citet{me-phd};
  (9)~\citet{2010A&A...521A..24J}; (10)~\citet{qk10};
  (11)~\citet{2008A&A...484..429S}; (12)~\citet{2006Natur.440..311S}.}

\end{deluxetable}

\clearpage
\begin{deluxetable}{lcccccc}
\tabletypesize{\scriptsize}
\tablecaption{Results of KS Tests \label{tbl:ks}}
\tablewidth{0pt}
\tablehead{
\colhead{} &
\multicolumn{6}{c}{Very Low-Mass Binaries}\\
\cline{2-7}
\colhead{Comparison Sample} &
\multicolumn{2}{c}{All} &
\multicolumn{2}{c}{Visual only} &
\multicolumn{2}{c}{$e<0.6$} \\
\colhead{} &
\colhead{Prob.} &
\colhead{$D$} &
\colhead{Prob.} &
\colhead{$D$} &
\colhead{Prob.} &
\colhead{$D$}}
\startdata

Solar-type binaries ($P > \Pcirc$)             & 0.50               & $0.23\pm0.04$ & 0.44               & $0.28\pm0.04$ & 0.57               & $0.23\pm0.04$ \\
Solar-type binaries ($\Pcirc < P < 10^3$~days) & 0.75               & $0.22\pm0.02$ & 0.82               & $0.22\pm0.04$ & 0.74               & $0.23\pm0.03$ \\
Solar-type binaries ($P > 10^3$~days)          & 0.25               & $0.31\pm0.05$ & 0.22               & $0.36\pm0.04$ & 0.46               & $0.28\pm0.03$ \\
$e^2$ \citep{ambartsumian37}                   & $9.1\times10^{-5}$ & $0.55\pm0.04$ & $2.2\times10^{-3}$ & $0.52\pm0.04$ & $5.3\times10^{-7}$ & $0.71\pm0.03$ \\
\citet{2009MNRAS.392..590B}                    & 0.83               & $0.20\pm0.04$ & 0.92               & $0.19\pm0.03$ & 0.64               & $0.26\pm0.04$ \\
\citet[][$a \leq 3.7$~AU]{2009MNRAS.392..590B} & 0.95               & $0.22\pm0.04$ & 0.97               & $0.22\pm0.03$ & 0.91               & $0.25\pm0.05$ \\
\citet{2009MNRAS.392..413S}                    & $9.9\times10^{-3}$ & $0.59\pm0.05$ & 0.034              & $0.56\pm0.05$ & 0.19               & $0.52\pm0.04$ \\
\citet[][$a \leq 3.7$~AU]{2009MNRAS.392..413S} & 0.028              & $0.56\pm0.05$ & 0.064              & $0.53\pm0.05$ & 0.43               & $0.45\pm0.05$ \\

\enddata

\tablecomments{Each table entry shows the probability that our sample
  of very low-mass orbits was drawn from the same parent distribution
  as the comparison sample along with the $D$ statistic and its rms
  scatter over $10^4$ Monte Carlo trials.}

\end{deluxetable}

\clearpage
\begin{deluxetable}{lccc}
\tabletypesize{\scriptsize}
\tablecaption{Conversion Factors from Projected Separation to Semimajor Axis \label{tbl:sma-corr}}
\tablewidth{0pt}
\tablehead{
\colhead{Assumed $e$ Distribution} &
\multicolumn{3}{c}{Correction factor ($a/\rho$)}\\
\cline{2-4}
\colhead{} &
\colhead{Median} &
\colhead{68.3\% c.l.} &
\colhead{95.4\% c.l.}}
\startdata

\multicolumn{4}{c}{\bf No discovery bias} \\
\cline{1-4}
Uniform ($0 < e < 1$)                          &  1.10  &  0.75, 2.02  &  0.57, 5.53  \\
Circular ($e=0$)                               &  1.15  &  1.01, 1.85  &  1.00, 4.72  \\
$e^2$ \citep{ambartsumian37}                   &  1.02  &  0.67, 2.07  &  0.55, 6.00  \\
Solar-type binaries ($P > \Pcirc$)             &  1.15  &  0.81, 2.01  &  0.64, 5.11  \\
Solar-type binaries ($\Pcirc < P < 10^3$~days) &  1.16  &  0.83, 2.00  &  0.66, 5.04  \\
Solar-type binaries ($P > 10^3$~days)          &  1.14  &  0.78, 2.02  &  0.61, 5.29  \\
Very low-mass (all binaries)                   &  1.16  &  0.84, 1.97  &  0.66, 5.08  \\
Very low-mass (visual binaries)                &  1.16  &  0.85, 1.97  &  0.66, 5.11  \\
\cline{1-4}

\multicolumn{4}{c}{} \\
\multicolumn{4}{c}{\bf Moderate discovery bias (IWA = $a/2$)} \\
\cline{1-4}
Uniform ($0 < e < 1$)                          &  1.02  &  0.72, 1.46  &  0.57, 1.89  \\
Circular ($e=0$)                               &  1.11  &  1.01, 1.46  &  1.00, 1.88  \\
$e^2$ \citep{ambartsumian37}                   &  0.91  &  0.65, 1.42  &  0.55, 1.88  \\
Solar-type binaries ($P > \Pcirc$)             &  1.06  &  0.79, 1.51  &  0.63, 1.90  \\
Solar-type binaries ($\Pcirc < P < 10^3$~days) &  1.07  &  0.80, 1.51  &  0.65, 1.90  \\
Solar-type binaries ($P > 10^3$~days)          &  1.04  &  0.76, 1.50  &  0.60, 1.90  \\
Very low-mass (all binaries)                   &  1.08  &  0.82, 1.51  &  0.66, 1.90  \\
Very low-mass (visual binaries)                &  1.08  &  0.82, 1.50  &  0.64, 1.90  \\
\cline{1-4}

\multicolumn{4}{c}{} \\
\multicolumn{4}{c}{\bf Severe discovery bias (IWA = $a$)\tablenotemark{a}} \\
\cline{1-4}
Uniform ($0 < e < 1$)                          &  0.79  &  0.63, 0.94  &  0.54, 0.99  \\
$e^2$ \citep{ambartsumian37}                   &  0.74  &  0.61, 0.91  &  0.53, 0.99  \\
Solar-type binaries ($P > \Pcirc$)             &  0.83  &  0.70, 0.95  &  0.60, 0.99  \\
Solar-type binaries ($\Pcirc < P < 10^3$~days) &  0.84  &  0.71, 0.95  &  0.61, 0.99  \\
Solar-type binaries ($P > 10^3$~days)          &  0.82  &  0.68, 0.94  &  0.58, 0.99  \\
Very low-mass (all binaries)                   &  0.84  &  0.72, 0.96  &  0.58, 0.99  \\
Very low-mass (visual binaries)                &  0.85  &  0.71, 0.96  &  0.58, 0.99  \\

\enddata

\tablenotetext{a}{When IWA $\geq a$, circular orbits are never detectable.}

\end{deluxetable}


\begin{thebibliography}{73}
\expandafter\ifx\csname natexlab\endcsname\relax\def\natexlab#1{#1}\fi

\bibitem[{Abt \& {Levy}(1976)}]{1976ApJS...30..273A}
Abt, H.~A., \& {Levy}, S.~G. 1976, \apjs, 30, 273

\bibitem[{Aitken(1918)}]{1918QB821.A3a......}
Aitken, R.~G. 1918, {The binary stars}, ed. {Aitken, R.~G.}

\bibitem[{Allen(2007)}]{2007ApJ...668..492A}
Allen, P.~R. 2007, \apj, 668, 492

\bibitem[{Ambartsumian(1937)}]{ambartsumian37}
Ambartsumian, V.~A. 1937, Astron.\ Zhurn., 14, 207

\bibitem[{Baraffe {et~al.}(2003)Baraffe, {Chabrier}, {Barman}, {Allard}, \&
  {Hauschildt}}]{2003A&A...402..701B}
Baraffe, I., {Chabrier}, G., {Barman}, T.~S., {Allard}, F., \& {Hauschildt},
  P.~H. 2003, \aap, 402, 701

\bibitem[{Basri \& {Mart{\'{\i}}n}(1999)}]{1999AJ....118.2460B}
Basri, G., \& {Mart{\'{\i}}n}, E.~L. 1999, \aj, 118, 2460

\bibitem[{Bate(2009{\natexlab{a}})}]{2009MNRAS.392..590B}
Bate, M.~R. 2009{\natexlab{a}}, \mnras, 392, 590

\bibitem[{Bate(2009{\natexlab{b}})}]{2009MNRAS.392.1363B}
---. 2009{\natexlab{b}}, \mnras, 392, 1363

\bibitem[{Bate(2010)}]{2010HiA....15..769B}
---. 2010, Highlights of Astronomy, 15, 769

\bibitem[{Berger {et~al.}(2011)Berger, {Monnier}, {Millan-Gabet}, {Renard},
  {Pedretti}, {Traub}, {Bechet}, {Benisty}, {Carleton}, {Haguenauer}, {Kern},
  {Labeye}, {Longa}, {Lacasse}, {Malbet}, {Perraut}, {Ragland}, {Schloerb},
  {Schuller}, \& {Thi{\'e}baut}}]{2011arXiv1103.3888B}
Berger, J., {et~al.} 2011, \aap, 529, L1

\bibitem[{Blake {et~al.}(2010)Blake, {Charbonneau}, \&
  {White}}]{2010ApJ...723..684B}
Blake, C.~H., {Charbonneau}, D., \& {White}, R.~J. 2010, \apj, 723, 684

\bibitem[{Blake {et~al.}(2008)Blake, {Charbonneau}, {White}, {Torres},
  {Marley}, \& {Saumon}}]{2008ApJ...678L.125B}
Blake, C.~H., {Charbonneau}, D., {White}, R.~J., {Torres}, G., {Marley}, M.~S.,
  \& {Saumon}, D. 2008, \apjl, 678, L125

\bibitem[{Bouy {et~al.}(2003)Bouy, {Brandner}, {Mart{\'{\i}}n}, {Delfosse},
  {Allard}, \& {Basri}}]{2003AJ....126.1526B}
Bouy, H., {Brandner}, W., {Mart{\'{\i}}n}, E.~L., {Delfosse}, X., {Allard}, F.,
  \& {Basri}, G. 2003, \aj, 126, 1526

\bibitem[{Bouy {et~al.}(2008)Bouy, {Mart{\'{\i}}n}, {Brandner}, {Forveille},
  {Delfosse}, {Hu{\'e}lamo}, {Basri}, {Girard}, {Zapatero Osorio}, {Stumpf},
  {Ghez}, {Valdivielso}, {Marchis}, {Burgasser}, \&
  {Cruz}}]{2008A&A...481..757B}
Bouy, H., {et~al.} 2008, \aap, 481, 757

\bibitem[{Brown(2004)}]{2004ApJ...607.1003B}
Brown, R.~A. 2004, \apj, 607, 1003

\bibitem[{Burgasser \& {Blake}(2009)}]{2009AJ....137.4621B}
Burgasser, A.~J., \& {Blake}, C.~H. 2009, \aj, 137, 4621

\bibitem[{Burgasser {et~al.}(2006)Burgasser, {Kirkpatrick}, {Cruz}, {Reid},
  {Leggett}, {Liebert}, {Burrows}, \& {Brown}}]{2006ApJS..166..585B}
Burgasser, A.~J., {Kirkpatrick}, J.~D., {Cruz}, K.~L., {Reid}, I.~N.,
  {Leggett}, S.~K., {Liebert}, J., {Burrows}, A., \& {Brown}, M.~E. 2006,
  \apjs, 166, 585

\bibitem[{Burgasser {et~al.}(2003)Burgasser, {Kirkpatrick}, {Reid}, {Brown},
  {Miskey}, \& {Gizis}}]{2003ApJ...586..512B}
Burgasser, A.~J., {Kirkpatrick}, J.~D., {Reid}, I.~N., {Brown}, M.~E.,
  {Miskey}, C.~L., \& {Gizis}, J.~E. 2003, \apj, 586, 512

\bibitem[{Burgasser {et~al.}(2008)Burgasser, {Liu}, {Ireland}, {Cruz}, \&
  {Dupuy}}]{2008ApJ...681..579B}
Burgasser, A.~J., {Liu}, M.~C., {Ireland}, M.~J., {Cruz}, K.~L., \& {Dupuy},
  T.~J. 2008, \apj, 681, 579

\bibitem[{Burgasser {et~al.}(2007)Burgasser, {Reid}, {Siegler}, {Close},
  {Allen}, {Lowrance}, \& {Gizis}}]{2007prpl.conf..427B}
Burgasser, A.~J., {Reid}, I.~N., {Siegler}, N., {Close}, L., {Allen}, P.,
  {Lowrance}, P., \& {Gizis}, J. 2007, in Protostars and Planets V, ed.
  B.~{Reipurth}, D.~{Jewitt}, \& K.~{Keil}, 427--441

\bibitem[{Burke(2008)}]{2008ApJ...679.1566B}
Burke, C.~J. 2008, \apj, 679, 1566

\bibitem[{Burrows {et~al.}(2001)Burrows, {Hubbard}, {Lunine}, \&
  {Liebert}}]{bur01}
Burrows, A., {Hubbard}, W.~B., {Lunine}, J.~I., \& {Liebert}, J. 2001, Reviews
  of Modern Physics, 73, 719

\bibitem[{Cardoso {et~al.}(2009)Cardoso, {McCaughrean}, {King}, {Close},
  {Scholz}, {Lenzen}, {Brandner}, {Lodieu}, \&
  {Zinnecker}}]{2009AIPC.1094..509C}
Cardoso, C.~V., {et~al.} 2009, in American Institute of Physics Conference
  Series, Vol. 1094, American Institute of Physics Conference Series, ed.
  {E.~Stempels}, 509--512

\bibitem[{Chabrier {et~al.}(2000)Chabrier, {Baraffe}, {Allard}, \&
  {Hauschildt}}]{2000ApJ...542..464C}
Chabrier, G., {Baraffe}, I., {Allard}, F., \& {Hauschildt}, P. 2000, \apj, 542,
  464

\bibitem[{Close {et~al.}(2003)Close, {Siegler}, {Freed}, \&
  {Biller}}]{2003ApJ...587..407C}
Close, L.~M., {Siegler}, N., {Freed}, M., \& {Biller}, B. 2003, \apj, 587, 407

\bibitem[{Close {et~al.}(2002)Close, {Siegler}, {Potter}, {Brandner}, \&
  {Liebert}}]{2002ApJ...567L..53C}
Close, L.~M., {Siegler}, N., {Potter}, D., {Brandner}, W., \& {Liebert}, J.
  2002, \apjl, 567, L53

\bibitem[{Costa {et~al.}(2005)Costa, {M{\'e}ndez}, {Jao}, {Henry},
  {Subasavage}, {Brown}, {Ianna}, \& {Bartlett}}]{2005AJ....130..337C}
Costa, E., {M{\'e}ndez}, R.~A., {Jao}, W.-C., {Henry}, T.~J., {Subasavage},
  J.~P., {Brown}, M.~A., {Ianna}, P.~A., \& {Bartlett}, J. 2005, \aj, 130, 337

\bibitem[{Costa {et~al.}(2006)Costa, {M{\'e}ndez}, {Jao}, {Henry},
  {Subasavage}, \& {Ianna}}]{2006AJ....132.1234C}
Costa, E., {M{\'e}ndez}, R.~A., {Jao}, W.-C., {Henry}, T.~J., {Subasavage},
  J.~P., \& {Ianna}, P.~A. 2006, \aj, 132, 1234

\bibitem[{Cruz {et~al.}(2007)Cruz, {Reid}, {Kirkpatrick}, {Burgasser},
  {Liebert}, {Solomon}, {Schmidt}, {Allen}, {Hawley}, \&
  {Covey}}]{2007AJ....133..439C}
Cruz, K.~L., {et~al.} 2007, \aj, 133, 439

\bibitem[{Dahn {et~al.}(2008)Dahn, {Harris}, {Levine}, {Tilleman}, {Monet},
  {Stone}, {Guetter}, {Canzian}, {Pier}, {Hartkopf}, {Liebert}, \&
  {Cushing}}]{2008ApJ...686..548D}
Dahn, C.~C., {et~al.} 2008, \apj, 686, 548

\bibitem[{Dahn {et~al.}(2002)}]{2002AJ....124.1170D}
---. 2002, \aj, 124, 1170

\bibitem[{Dupuy(2010)}]{me-phd}
Dupuy, T.~J. 2010, PhD thesis, University of Hawai'i at Manoa

\bibitem[{Dupuy {et~al.}(2009{\natexlab{a}})Dupuy, {Liu}, \&
  {Bowler}}]{me-2206}
Dupuy, T.~J., {Liu}, M.~C., \& {Bowler}, B.~P. 2009{\natexlab{a}}, \apj, 706,
  328

\bibitem[{Dupuy {et~al.}(2010)Dupuy, {Liu}, {Bowler}, \& {Cushing}}]{me-latem}
Dupuy, T.~J., {Liu}, M.~C., {Bowler}, B.~P., \& {Cushing}, M.~C. 2010, ApJ,
  721, 1725

\bibitem[{Dupuy {et~al.}(2009{\natexlab{b}})Dupuy, {Liu}, \&
  {Ireland}}]{me-130948}
Dupuy, T.~J., {Liu}, M.~C., \& {Ireland}, M.~J. 2009{\natexlab{b}}, \apj, 692,
  729

\bibitem[{Dupuy {et~al.}(2009{\natexlab{c}})Dupuy, {Liu}, \&
  {Ireland}}]{me-2397a}
---. 2009{\natexlab{c}}, \apj, 699, 168

\bibitem[{Duquennoy \& {Mayor}(1991)}]{1991A&A...248..485D}
Duquennoy, A., \& {Mayor}, M. 1991, \aap, 248, 485

\bibitem[{Fischer \& {Marcy}(1992)}]{1992ApJ...396..178F}
Fischer, D.~A., \& {Marcy}, G.~W. 1992, \apj, 396, 178

\bibitem[{Gatewood \& {Coban}(2009)}]{2009AJ....137..402G}
Gatewood, G., \& {Coban}, L. 2009, \aj, 137, 402

\bibitem[{Ghez {et~al.}(2008)Ghez, {Salim}, {Weinberg}, {Lu}, {Do}, {Dunn},
  {Matthews}, {Morris}, {Yelda}, {Becklin}, {Kremenek}, {Milosavljevic}, \&
  {Naiman}}]{2008ApJ...689.1044G}
Ghez, A.~M., {et~al.} 2008, \apj, 689, 1044

\bibitem[{Gizis {et~al.}(2000)Gizis, {Monet}, {Reid}, {Kirkpatrick}, {Liebert},
  \& {Williams}}]{2000AJ....120.1085G}
Gizis, J.~E., {Monet}, D.~G., {Reid}, I.~N., {Kirkpatrick}, J.~D., {Liebert},
  J., \& {Williams}, R.~J. 2000, \aj, 120, 1085

\bibitem[{Harrington \& {Miranian}(1977)}]{1977PASP...89..400H}
Harrington, R.~S., \& {Miranian}, M. 1977, \pasp, 89, 400

\bibitem[{Joergens \& {M{\"u}ller}(2007)}]{2007ApJ...666L.113J}
Joergens, V., \& {M{\"u}ller}, A. 2007, \apjl, 666, L113

\bibitem[{Joergens {et~al.}(2010)Joergens, {M{\"u}ller}, \&
  {Reffert}}]{2010A&A...521A..24J}
Joergens, V., {M{\"u}ller}, A., \& {Reffert}, S. 2010, \aap, 521, A24

\bibitem[{Kirkpatrick {et~al.}(2000)Kirkpatrick, {Reid}, {Liebert}, {Gizis},
  {Burgasser}, {Monet}, {Dahn}, {Nelson}, \& {Williams}}]{2000AJ....120..447K}
Kirkpatrick, J.~D., {et~al.} 2000, \aj, 120, 447

\bibitem[{Konopacky {et~al.}(2010)Konopacky, {Ghez}, {Barman}, {Rice},
  {Bailey}, {White}, {McLean}, \& {Duch{\^e}ne}}]{qk10}
Konopacky, Q.~M., {Ghez}, A.~M., {Barman}, T.~S., {Rice}, E.~L., {Bailey},
  J.~I., {White}, R.~J., {McLean}, I.~S., \& {Duch{\^e}ne}, G. 2010, \apj, 711,
  1087

\bibitem[{Kraus \& {Hillenbrand}(2009)}]{2009ApJ...703.1511K}
Kraus, A.~L., \& {Hillenbrand}, L.~A. 2009, \apj, 703, 1511

\bibitem[{L{\'e}pine {et~al.}(2009)L{\'e}pine, {Thorstensen}, {Shara}, \&
  {Rich}}]{2009AJ....137.4109L}
L{\'e}pine, S., {Thorstensen}, J.~R., {Shara}, M.~M., \& {Rich}, R.~M. 2009,
  \aj, 137, 4109

\bibitem[{Liu {et~al.}(2008)Liu, {Dupuy}, \& {Ireland}}]{liu-2m1534}
Liu, M.~C., {Dupuy}, T.~J., \& {Ireland}, M.~J. 2008, \apj, 689, 436

\bibitem[{Liu {et~al.}(2010)Liu, {Dupuy}, \& {Leggett}}]{liu-1209}
Liu, M.~C., {Dupuy}, T.~J., \& {Leggett}, S.~K. 2010, \apj, 722, 311

\bibitem[{Mathieu(1994)}]{1994ARA&A..32..465M}
Mathieu, R.~D. 1994, \araa, 32, 465

\bibitem[{Monet {et~al.}(1992)Monet, {Dahn}, {Vrba}, {Harris}, {Pier},
  {Luginbuhl}, \& {Ables}}]{1992AJ....103..638M}
Monet, D.~G., {Dahn}, C.~C., {Vrba}, F.~J., {Harris}, H.~C., {Pier}, J.~R.,
  {Luginbuhl}, C.~B., \& {Ables}, H.~D. 1992, \aj, 103, 638

\bibitem[{Pourbaix {et~al.}(2004)Pourbaix, {Tokovinin}, {Batten}, {Fekel},
  {Hartkopf}, {Levato}, {Morrell}, {Torres}, \& {Udry}}]{2004A&A...424..727P}
Pourbaix, D., {et~al.} 2004, \aap, 424, 727

\bibitem[{Raghavan {et~al.}(2010)Raghavan, {McAlister}, {Henry}, {Latham},
  {Marcy}, {Mason}, {Gies}, {White}, \& {ten Brummelaar}}]{2010ApJS..190....1R}
Raghavan, D., {et~al.} 2010, \apjs, 190, 1

\bibitem[{Reid {et~al.}(2001)Reid, {Gizis}, {Kirkpatrick}, \&
  {Koerner}}]{2001AJ....121..489R}
Reid, I.~N., {Gizis}, J.~E., {Kirkpatrick}, J.~D., \& {Koerner}, D.~W. 2001,
  \aj, 121, 489

\bibitem[{Reid {et~al.}(2006)Reid, {Lewitus}, {Allen}, {Cruz}, \&
  {Burgasser}}]{2006AJ....132..891R}
Reid, I.~N., {Lewitus}, E., {Allen}, P.~R., {Cruz}, K.~L., \& {Burgasser},
  A.~J. 2006, \aj, 132, 891

\bibitem[{Reipurth \& {Clarke}(2001)}]{2001AJ....122..432R}
Reipurth, B., \& {Clarke}, C. 2001, \aj, 122, 432

\bibitem[{Reipurth {et~al.}(2007)Reipurth, {Guimar{\~a}es}, {Connelley}, \&
  {Bally}}]{2007AJ....134.2272R}
Reipurth, B., {Guimar{\~a}es}, M.~M., {Connelley}, M.~S., \& {Bally}, J. 2007,
  \aj, 134, 2272

\bibitem[{Rice {et~al.}(2003)Rice, {Armitage}, {Bonnell}, {Bate}, {Jeffers}, \&
  {Vine}}]{2003MNRAS.346L..36R}
Rice, W.~K.~M., {Armitage}, P.~J., {Bonnell}, I.~A., {Bate}, M.~R., {Jeffers},
  S.~V., \& {Vine}, S.~G. 2003, \mnras, 346, L36

\bibitem[{Schaefer {et~al.}(2006)Schaefer, {Simon}, {Beck}, {Nelan}, \&
  {Prato}}]{2006AJ....132.2618S}
Schaefer, G.~H., {Simon}, M., {Beck}, T.~L., {Nelan}, E., \& {Prato}, L. 2006,
  \aj, 132, 2618

\bibitem[{Seifahrt {et~al.}(2008)Seifahrt, {R{\"o}ll}, {Neuh{\"a}user},
  {Reiners}, {Kerber}, {K{\"a}ufl}, {Siebenmorgen}, \&
  {Smette}}]{2008A&A...484..429S}
Seifahrt, A., {R{\"o}ll}, T., {Neuh{\"a}user}, R., {Reiners}, A., {Kerber}, F.,
  {K{\"a}ufl}, H.~U., {Siebenmorgen}, R., \& {Smette}, A. 2008, \aap, 484, 429

\bibitem[{Shen \& {Turner}(2008)}]{2008ApJ...685..553S}
Shen, Y., \& {Turner}, E.~L. 2008, \apj, 685, 553

\bibitem[{Siegler {et~al.}(2007)Siegler, {Close}, {Burgasser}, {Cruz},
  {Marois}, {Macintosh}, \& {Barman}}]{2007AJ....133.2320S}
Siegler, N., {Close}, L.~M., {Burgasser}, A.~J., {Cruz}, K.~L., {Marois}, C.,
  {Macintosh}, B., \& {Barman}, T. 2007, \aj, 133, 2320

\bibitem[{Siegler {et~al.}(2003)Siegler, {Close}, {Mamajek}, \&
  {Freed}}]{2003ApJ...598.1265S}
Siegler, N., {Close}, L.~M., {Mamajek}, E.~E., \& {Freed}, M. 2003, \apj, 598,
  1265

\bibitem[{Stamatellos {et~al.}(2007)Stamatellos, {Hubber}, \&
  {Whitworth}}]{2007MNRAS.382L..30S}
Stamatellos, D., {Hubber}, D.~A., \& {Whitworth}, A.~P. 2007, \mnras, 382, L30

\bibitem[{Stamatellos \& {Whitworth}(2009)}]{2009MNRAS.392..413S}
Stamatellos, D., \& {Whitworth}, A.~P. 2009, \mnras, 392, 413

\bibitem[{Stassun {et~al.}(2006)Stassun, {Mathieu}, \&
  {Valenti}}]{2006Natur.440..311S}
Stassun, K.~G., {Mathieu}, R.~D., \& {Valenti}, J.~A. 2006, \nat, 440, 311

\bibitem[{Tinney {et~al.}(2003)Tinney, {Burgasser}, \&
  {Kirkpatrick}}]{2003AJ....126..975T}
Tinney, C.~G., {Burgasser}, A.~J., \& {Kirkpatrick}, J.~D. 2003, \aj, 126, 975

\bibitem[{Torres(1999)}]{1999PASP..111..169T}
Torres, G. 1999, \pasp, 111, 169

\bibitem[{van Leeuwen(2007)}]{2007hnrr.book.....V}
van Leeuwen, F. 2007, {Hipparcos, the New Reduction of the Raw Data}
  (Hipparcos, the New Reduction of the Raw Data.~By Floor van Leeuwen,
  Institute of Astronomy, Cambridge University, Cambridge, UK Series:
  Astrophysics and Space Science Library, Vol.~ 350 20 Springer Dordrecht)

\bibitem[{Vrba {et~al.}(2004)}]{2004AJ....127.2948V}
Vrba, F.~J., {et~al.} 2004, \aj, 127, 2948

\bibitem[{Worley \& {Heintz}(1983)}]{1983PUSNO..24g...1W}
Worley, C.~E., \& {Heintz}, W.~D. 1983, Publications of the U.S.~Naval
  Observatory Second Series, 24, 1

\bibitem[{Zahn \& {Bouchet}(1989)}]{1989A&A...223..112Z}
Zahn, J., \& {Bouchet}, L. 1989, \aap, 223, 112

\end{thebibliography}
\end{document}